\documentclass[sigconf]{acmart}
\usepackage{longtable}
\usepackage{tabularx}
\usepackage{fontawesome}
\usepackage{pdfpages}
\usepackage{graphicx}
\usepackage{booktabs}
\usepackage{multirow}
\usepackage{multicol,lipsum}
\usepackage{xcolor}
\usepackage{dblfloatfix}  
\usepackage{soul}
\usepackage{pdflscape}
\usepackage{supertabular,booktabs}

\newif\ifshowchanges
\showchangestrue   

\AtBeginDocument{%
  \providecommand\BibTeX{{%
    \normalfont B\kern-0.5em{\scshape i\kern-0.25em b}\kern-0.8em\TeX}}}
\setcopyright{acmcopyright}
\copyrightyear{2024}
\acmYear{2024}
\setcopyright{acmlicensed}\acmConference[DIS '24]{Designing Interactive Systems Conference}{July 1--5, 2024}{IT University of Copenhagen, Denmark}
\acmBooktitle{Designing Interactive Systems Conference (DIS '24), July 1--5, 2024, IT University of Copenhagen, Denmark}
\acmDOI{10.1145/3643834.3661587}
\acmISBN{979-8-4007-0583-0/24/07}

\begin{document}


\title[Not Just Novelty: A Longitudinal Study on Utility and Customization of an AI Workflow]{Not Just Novelty: A Longitudinal Study on Utility and Customization of an AI Workflow}

\author{Tao Long}
\email{long@cs.columbia.edu}
\affiliation{%
  \institution{Columbia University}
  \city{New York City}
  \state{New York}
  \country{USA}
}
\author{Katy Ilonka Gero}
\email{katy@g.harvard.edu}
\affiliation{%
  \institution{Harvard University}
  \city{Cambridge}
  \state{Massachusetts}
  \country{USA}
}

\author{Lydia B. Chilton}
\email{chilton@cs.columbia.edu}
\affiliation{%
  \institution{Columbia University}
  \city{New York City}
  \state{New York}
  \country{USA}
}

\renewcommand{\shortauthors}{Tao Long, Katy Ilonka Gero, and Lydia B. Chilton}

\begin{abstract}
Generative AI brings novel and impressive abilities to help people in everyday tasks. There are many AI workflows that solve real and complex problems by chaining AI outputs together with human interaction. Although there is an undeniable lure of AI, it is uncertain how useful generative AI workflows are after the novelty wears off. Additionally, workflows built with generative AI have the potential to be easily customized to fit users' individual needs, but do users take advantage of this? We conducted a three-week longitudinal study with 12 users to understand the familiarization and customization of generative AI tools for science communication. Our study revealed that there exists a familiarization phase, during which users were exploring the novel capabilities of the workflow and discovering which aspects they found useful. After this phase, users understood the workflow and were able to anticipate the outputs. Surprisingly, after familiarization the perceived utility of the system was rated higher than before, indicating that the perceived utility of AI is not just a novelty effect. The increase in benefits mainly comes from end-users' ability to customize prompts, and thus potentially appropriate the system to their own needs. This points to a future where generative AI systems can allow us to design for appropriation.




\end{abstract}




\maketitle

\section{Introduction}

Generative AI has impressive abilities to generate text, images, and code based on simple prompts. Building on this, a powerful technique to solve more complex problems is to break problems into AI workflows that chain together prompts~\cite{ai_chains}. 
Often, the steps of the chain involve interfaces with human interaction to check the results and steer progress in productive directions. 
This can enable more complex processes such as following divergent and convergent thinking \cite{popblends,graphologue}, the human-centered design process \cite{opal,anglekindling,ReelFramer}, the process model of writing~\cite{mirowski2022cowriting,Sparks, visar}, and simulating thoughts, behaviors and actions of a group of people interacting~\cite{agent}. 
Studies of AI workflows tend to show that people find them useful. They save time~\cite{Tweetorial_ICCC} and mental effort~\cite{anglekindling,mirowski2022cowriting}, and increase output~\cite {opal}.

However, most of these systems test their utility in a single lab session. 
The presence of AI in a workflow brings great excitement, expectations, and hype. But could this just be an effect of the novelty of including AI in a workflow?
For workflows that provide scaffolding, it is natural to assume that scaffolding might not be needed after the user has internalized the workflow. Workflows are often criticized for pushing users towards similar outputs and constraining diversity and creativity. Moreover, workflows can be brittle \cite{workflowbrittle}--- although they may work for one set of well-defined tasks, they might not generalize to more complex examples that users encounter in the wild. Is the success of AI workflows a product of the novelty effect, or is there a lasting benefit?

The novelty effect is a common effect seen in technology adoption~\cite{rodrigues_gamification_2022, novelty_microsoft, fitbit_novelty} where the desire to use a new technology is initially high but then diminishes over time --- as users become more familiar with the technology it seems less magical.
Previous studies show that some technologies overcome the novelty effect (e.g., gamified learning systems~\cite{rodrigues_gamification_2022,gamifiedstudy2}), but some do not (e.g., health activity trackers~\cite{fitbit_novelty} and smart speakers~\cite{10168160}). 
Even though there is a novelty effect, the familiarization effect can actually increase engagement over time ~\cite{rodrigues_gamification_2022}. Technology that becomes familiar can become internalized and thus used with less effort, leading to greater rewards. 
Long-term studies on education technologies indicate that technologies that cater to users' needs and feelings of autonomy can overcome the initial novelty effect~\cite {novelty_selfdetermination}. Currently, it is unclear whether the appeal of AI workflows stems from a novelty effect or if there is a lasting benefit past the workflow familiarization.

We conducted a three-week longitudinal study with 12 users to understand the familiarization and customization of generative AI tools for science communication. 
We study a 7-step human-AI workflow to help STEM experts write tweets that motivate a topic in their field to the general public~\cite{Tweetorial_ICCC}. 
The science communication task is called a Tweetorial hook~\cite{Tweetorial_CSCW}, which is challenging because it requires STEM experts to think about examples or experiences with the topic the public might relate to. For example, the web programming topic of AJAX is very technical, but users have all experienced it on social media when new posts load as they scroll without users explicitly requesting them. 
A previous study~\cite{Tweetorial_ICCC} shows that in a single session, a workflow based on a large language model (LLM) reduced the mental demand for writing a Tweetorial hook compared to writing without it (simply using web search).
We recreated this workflow and added two simple customization features. First, we display the LLM prompt used at every stage and allow users to edit it to customize the responses and the workflow. Second, we add the hooks written by the users to the prompts as training examples so the system can be personalized based on examples.
Then, we study a three-week deployment of the system as design probe~\cite{technologyprobe}, over which participants use the system in 10 different sessions to write ten different hooks. We identify if and when familiarization occurs. To understand whether there is a novelty effect, we use surveys, observations, and interviews to assess the system's usefulness before and after familiarization. 

\vspace{3px}
Our study revealed that:
\begin{itemize}
\item A familiarization phase exists and lasts for a few sessions (4.27 sessions, during which the system is already perceived as useful). 
\setlength\itemsep{2px}
\item Before the familiarization concluded, users were exploring the novel capabilities of the workflow and discovering which aspects they found useful.
\setlength\itemsep{2px}
\item After familiarization, the system is rated more useful, indicating that the novelty does not wear off (a 12.1\% improvement, p-value < 0.005). 
\setlength\itemsep{2px}
\item This change in the system's usefulness can mainly be attributed to the increased usefulness of the prompt editing feature (an 11.4\% improvement, p-value < 0.005). The other features show no change in usefulness throughout the study.
\setlength\itemsep{2px}
\item Users edit prompts for three major reasons: to keep generations relevant, to guide narrative directions and styles, and to make the workflow more efficient. 
Chaining users' previous hooks to the prompt as training examples shows no significant benefits.
\setlength\itemsep{2px}
\item {After familiarization, user's mental models of the workflow did not change, but their perceptions of ownership increased. There were nuanced ways that users experienced ownership after multiple experiences with the system, but there is a clear relationship between involvement and ownership. }
\vspace{2px}
\end{itemize}

We conclude by exploring why the novelty did not wear off and generalizing its implications for other AI workflows.

\section{Related Work}

\subsection{Generative AI-Based Workflows}

Generative AI models hold huge potential for supporting various tasks with their domain knowledge and contextual understanding capabilities~\cite{gpt4, bommasani_opportunities_2022, chatgptproductivity}. 
For example, users can use LLMs for simple tasks like answering math questions and complex ones such as reviewing and editing academic papers or even writing screenplay scripts~\cite{bommasani_opportunities_2022, mirowski2022cowriting}. 
However, LLMs have many limitations. 
First, LLMs face challenges in producing high-quality outputs when they have limited knowledge about the task~\cite{Eloundou_Manning_Mishkin_Rock_2023} or multi-step reasoning is required~\cite{chain_of_thought}. Second, due to the limitations of the underlying generative models, LLM outputs often contain hallucination, repetitiveness, and vague language~\cite{holtzman_curious_2020,ippolito_comparison_2019, bommasani_opportunities_2022}. Third, as LLMs can take any natural language prompts as inputs, users may find it challenging to determine how to engage and collaborate with the LLMs to troubleshoot their prompts for enhanced outputs~\cite{ai_chains}. 

AI workflows help solve complex real-life problems and mitigate LLM shortcomings. As the foundation structure of workflows, AI chains~
\cite{ai_chains} (or LLM chaining) break down complex tasks into simpler steps then chain the LLM output of one step to the input of the next LLM step~\cite{grunde_mclaughlin_designing_2023, PromptChainer, ai_chains}.  
This multi-step process helps complete complex tasks that LLMs are not able to do well in one prompt. 
Also, chaining allows users to easily evaluate the outputs within each step to identify errors, thus providing user control to find errors, fix them, and improve generations according to their preferences~\cite{grunde_mclaughlin_designing_2023, Tweetorial_ICCC, wu_llms_2023}. 

AI workflows successfully complement users with guided support in many traditionally cognitively demanding tasks. For example, various workflows demonstrate their strong system performances or satisfactory user experiences to support writing~\cite{ReelFramer, Tweetorial_ICCC, mirowski2022cowriting,li_teach_2023, schick_peer_2022, yang_re3_2022, petridis_constitutionmaker_2023, 10.1145/3584931.3607492}, brainstorming~\cite{popblends, anglekindling, bursztyn_learning_2022, kim_metaphorian_2023, wu_llms_2023, visar, choi_creativeconnect_2023, seo_chacha_2024, suh_structured_2023}, visual design~\cite{opal, disco, graphologue, promptinfuser, stylette, tseng_keyframer_2024, son_genquery_2023, feng_canvil_2024, xiao_typedance_2024, huang_graphimind_2024} and question answering and sensemaking ~\cite{eisenstein_honest_2022, Xie2023SelfEvaluationGB, du_improving_2023, kazemitabaar_codeaid_2024, chainforge, gao_collabcoder_2024, suh_sensecape_2023, zhou_instructpipe_2023, li_eliciting_2023}.

However, most of these studies are conducted solely within the framework of controlled user studies lasting typically 30 to 60 minutes. This approach might only capture users' initial reactions of satisfaction or dissatisfaction with the system rather than providing more comprehensive experiences ~\cite{userstudy_consideredharmful,blaynee2016collaborative, sporka2011chanti,jain2012user}. Thus, a gap exists in understanding how AI workflows perform and whether they continue to aid users well after repeated usage.

\subsection{Novelty Effect}
The novelty effect is a common effect seen in technology adoption~\cite{lai2017literature, koch2018novelty, mira}, where the desire to use new technology is initially high, but then diminishes over time. 
It is related to humans' innate desire for novelty and our psychological stress response: individuals exhibit a stronger stress response the first time they face a challenging or threatening experience~\cite{long9,hekkert_most_2003}. With the stress, users are more likely to focus, embrace learning through trials and errors with a higher tolerance, decrease expectation, and perceive higher performance~\cite{9288720,long9, placebo, nielsen_longitudinal_2021,miguel2024evaluation}. However, novelty effects will eventually diminish as users become more familiar with the technology, with the duration varying depending on the technology~\cite{koch2018novelty, lai2017literature}.

After the novelty wears off, some technologies fail to overcome this effect, wherein users find them less useful after becoming familiar with them, ultimately leading to dissatisfaction and abandonment
~\cite{novelty_microsoft,fitbit_novelty,10.1145/3428361.3428469,long_challenges_2023,10168160}. However, many technologies successfully navigate this challenge, as users continue to find them useful over time~\cite{novelty_selfdetermination, rodrigues_gamification_2022,gamifiedstudy2}. Also, even though the novelty effect plays a role, familiarizing these technologies can increase users' engagement and perceived usefulness over time~\cite{rodrigues_gamification_2022}. Technology that becomes familiar can become internalized and thus used with less effort, leading to greater rewards.
Additionally, long-term studies on education technologies indicate that tools that cater to users' needs and feelings of autonomy can overcome the initial novelty effect~\cite{novelty_selfdetermination}. This is aligned with self-determination theory~\cite{deci_self_determination} in educational tools and technology: when the needs for autonomy, competence, and relatedness are satisfied, positive outcomes follow~\cite{baumeister1995need,ryan2017self, 10.1145/97243.97271}. 

Generative AI workflows also draw concern regarding the strong novelty effect due to their seemingly dazzling performance~\cite{mukherjee_creative_2023}. There exists a gap in examining if the benefits of AI tools and workflows hold up and what happens after the novelty wears off.

\subsection{Technology Customization and Appropriation}
Users customize and appropriate the technology over usage and time. 
Many long-term studies~\cite{longitudinalReview, long1, long2, long4, long5, long6, long7, long8} and technology domestication theory~\cite{berker2005domestication} mapped user's journey of adopting and customizing these technologies. Beginning with the learning and comprehending the system affordances, especially the essential features, users navigate the learning phase ~\cite{mira, long3}. Once they better understand the technology and the supported task, users engage in customization to shape the technology for better alignment with their own preference~\cite{long7, TheoryOfAppropriation, moran_everyday_2002,soyagroup}. 
Then, users start to appropriate the technology by modifying it and applying it for new use cases that the original designers had not anticipated~\cite{DesigningForAppropriationAlanDix, carrollappropriation, dourish_appropriation_2003, mullerappro1, mullerappro2}. Along with the increased familiarization and insights gained over usages, users gradually develop their own perceptions, mental models, and opinions towards the system \cite{Ross, long7,socialdynamics, longitudinalReview}. Technologies that allow customization and appropriation can elevate user satisfaction and performance and meet more user needs~\cite{DesigningForAppropriationAlanDix, 10.1145/3457151, feng_canvil_2024,10.1145/3586183.3606833}. Both customization and appropriation offer users a better sense of control and ownership over the system and its generations~\cite{DesigningForAppropriationAlanDix, 10.1145/1978942.1979056}. 

A 2007 paper \cite{DesigningForAppropriationAlanDix} offers guidelines to design systems that support customization and appropriation, outlining three key principles. First, the system should "\textit{allow interpretation}": enable users to assign their interpretation to system elements instead of fixing them. Second, it needs to "\textit{provide visibility}": make the system's functionality transparent for users to understand the effects of their actions. Third, it needs to "\textit{support not control}" and offer "\textit{plugability and configuration}": design for efficiency and flexibility, allowing the system to automate and users to combine parts of the system according to their needs.

For generative AI workflows, natural language prompts largely simplify the customization process. Given the flexibility to alter the system through word-level changes, users can easily customize workflows by tweaking prompts or incorporating examples from their previous work.
It broadens the access to customization for non-experts~\cite{JohnnyPrompt, prompt_engineering}, in contrast to traditional methods requiring developers and designers to engage directly with users for system updates. While this form of customization holds the potential to enhance the utility of AI systems over time, it remains uncertain whether it can overcome the novelty effect and how users will actually engage in customization or appropriation~\cite{scaffoldinglevel}.

\section{Background on Tweetorial Hooks}

\subsection{Science Communication and Tweetorials}

Communicating science effectively and engagingly is key for the public to understand and navigate our rapidly changing world. Historically, the realm of science communication was largely confined within the boundaries of scholarly papers~\cite{Bucchi_Trench_2021}. These papers are intended for a niche audience of fellow scientists and scholars, and are often difficult for lay people to understand because the writing contains jargon, assumes background knowledge and does not motivate the material or make it relatable ~\cite{Williams_Jones_Reinecke_Hsieh_2022}. However, with the rise of social media, there has been a marked shift in how people access scientific information online: more and more people are starting to learn science in fragmented time on social media~\cite{hargittaiHowYoungAdults2018,Williams_Jones_Reinecke_Hsieh_2022}. 

Starting in late 2017, Tweetorials have become a popular science communication medium for STEM experts to reach the general audience~\cite{Symplur_2019}. A 2021 study analyzed popular Tweetorials to identify the science communication principles that made them effective~\cite{Tweetorial_CSCW}. They found that Tweetorials are generally short (threads of 8 - 15 tweets) and use informal, narrative-driven, and personal language.

\begin{figure}[!b]
\vspace{-15px}
 \begin{center}
\includegraphics[width=\linewidth]{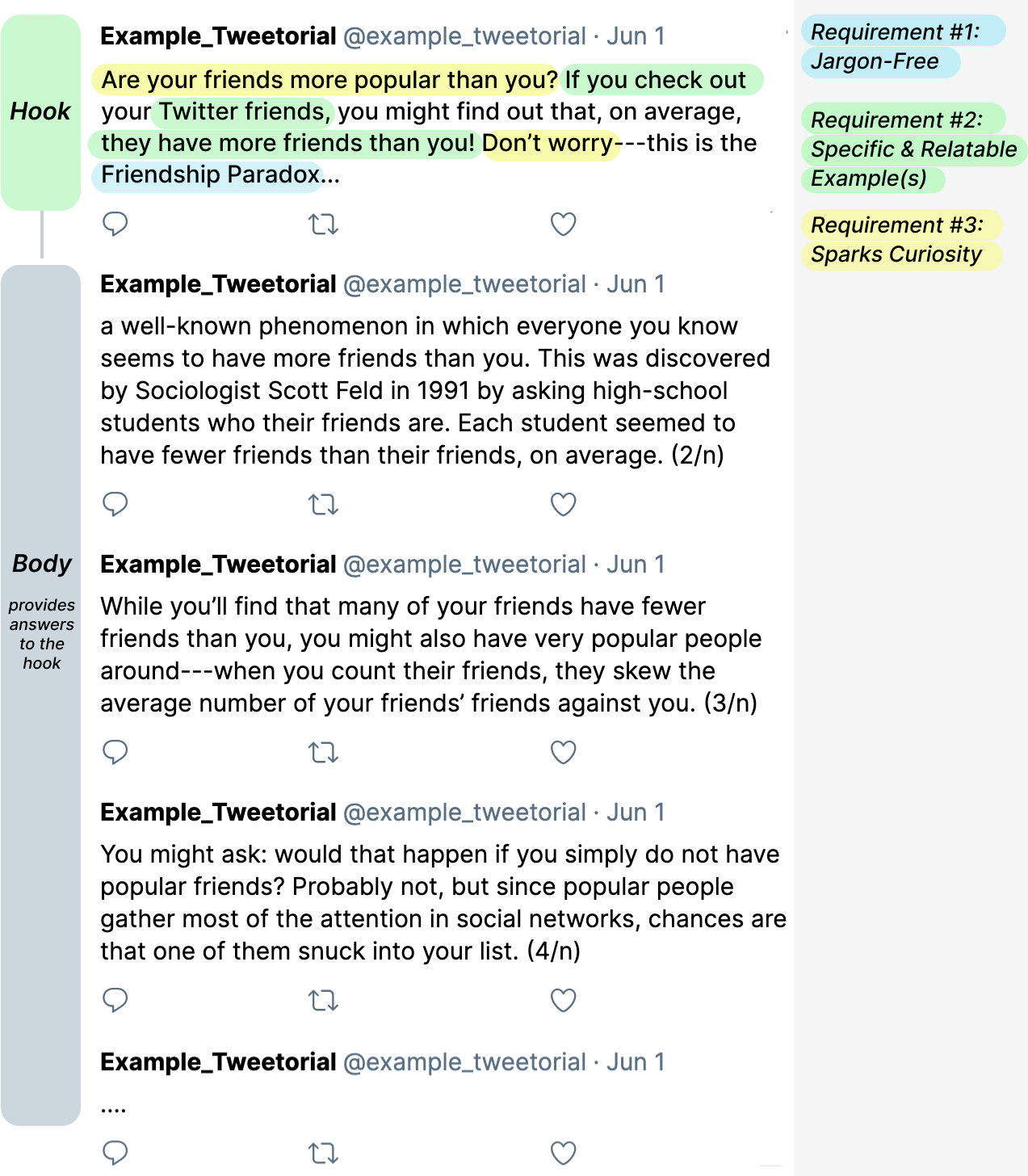}
\Description[A Tweetorial example about friendship paradox]{A Tweetorial example about the friendship paradox, a technical concept in social networks, annotated for its structure. It comes with a hook and later body sections. The hook is highlighted in green: "Are your friends more popular than you? If you check out your Twitter friends, you might find out that, on average, they have more friends than you! Don’t worry---this is the Friendship Paradox... (1/n)."}
\caption{An example Tweetorial about ``\textit{friendship paradox}", a technical concept in social networks, annotated for its structure and adherence to the requirements (See section 3.2). It comes with a hook and later body sections.}
\label{fig:tweetorial_example}
 \end{center}
 \vspace{-5px}

\end{figure}

As shown in Figure \ref{fig:tweetorial_example}, the key feature of a Tweetorial is the first tweet, which is called the ``hook'' to motivate the topic and draw readers in. The hook has an intriguing question: “\textit{Are your friends more popular than you?}” It provides a relatable idea: “\textit{If you check out your Twitter friends, you might find that on average they have more friends than you.}” This introduces the topic of the Friendship Paradox without jargon. And the details make it really personal as something the reader may have encountered in their own life. Then, the rest of the Tweetorial is the “body” that explains the concept, where the hook can pique the reader's interest in continuing to read about this real-world experience.

However, studies on Tweetorial writing have shown that writing hooks are a key challenge for STEM experts~\cite{Tweetorial_CSCW}. They are trained to write about their work in a formal tone for other experts, and it is difficult to go against that training. Another difficulty is finding examples and layman’s terms to motivate the topic for general audiences. Also, they feel uncomfortable using subjective and informal language and avoid personal details, even though 80\% of the Tweetorials have them. In an exploratory study using LLMs to support Tweetorial writers, one of the major use cases was ideating concrete examples for the hook~\cite{Sparks}. This indicates there is potential to use LLMs to help write Tweetorial hooks. 

\subsection{What Makes a Good Tweetorial Hook?}
Previous work~\cite{Tweetorial_CSCW} has analyzed Tweetorial hooks using science communication principles, and they found that most hooks had three important characteristics, which were subsequently formulated into requirements for assessing an effective hook by~\cite{Tweetorial_ICCC}: 

\begin{itemize}
\vspace{5px}
    \item \underline{Requirement 1: Jargon-Free.} The hook should contain no jargon so the general audience can understand it easily, with the possible exception of naming the topic to intrigue readers.
     \vspace{2px}
    \item \underline{Requirement 2: Specific and Relatable Example(s).} The hook should include an analogy, a current event, a popular misconception, or a common experience.
     \vspace{2px}
    \item \underline{Requirement 3: Sparks Curiosity.} The hook should include an intriguing question that sparks readers’ curiosity and drives readers to continue reading.
    \vspace{5px}
\end{itemize}

Figure \ref{fig:tweetorial_example_LLM} is an example of Tweetorial hooks for Language Models, a computer science topic, exhibiting these properties: 

\begin{figure}[h]
 \begin{center}
\includegraphics[width=\linewidth]{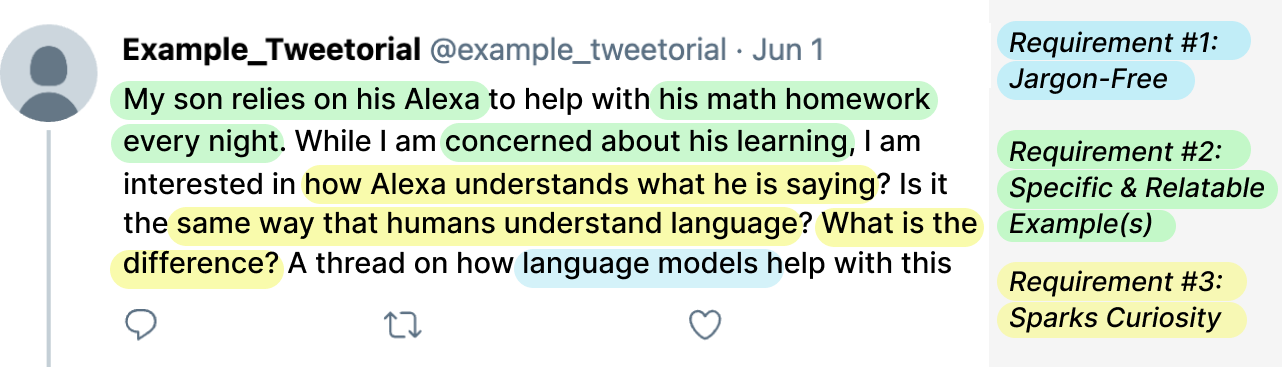}
\Description[A Tweetorial hook example about language models]{A Tweetorial hook about language models, a technical concept in computer science: "My son relies on his Alexa to help with his math homework every single night. While I am concerned about his learning, I am interested in how Alexa understands what he is saying? Is it the same way that humans understand language? What is the difference? A thread on how language models help with this:" Then, the example hook is annotated for its adherence to the requirements. In this hook, the relatable experience is talking to a smart device, like Alexa. The specific details include parents' concerns about the kids' learning and Alexa helping with math homework. The concerns about his son spark curiosity: “how Alexa understand what he is saying?”}
\caption{An example Tweetorial hook about ``\textit{language models}", a technical concept in computer science, annotated for its adherence to the requirements. In this hook, the relatable experience is talking to a smart device at home, like Amazon Alexa. The specific details include parents' concerns about the kids' learning and Alexa helping with math homework. The concerns about his son spark curiosity: “\textit{how Alexa understand what he is saying?}”}
\label{fig:tweetorial_example_LLM}
\end{center}
\vspace{-5px}
\end{figure}

As shown, the relatable and specific content is an experience with the technology. Although this doesn’t work for all topics, it tends to work well for computer science topics because those concepts are usually the technologies we all rely on and can be comparatively smoothly contextualized into real-life use cases. 

\subsection{Existing AI Workflow for Tweetorial Hooks}

\begin{figure}[h]
 \begin{center}
\vspace{-5px}
\includegraphics[width=\linewidth]{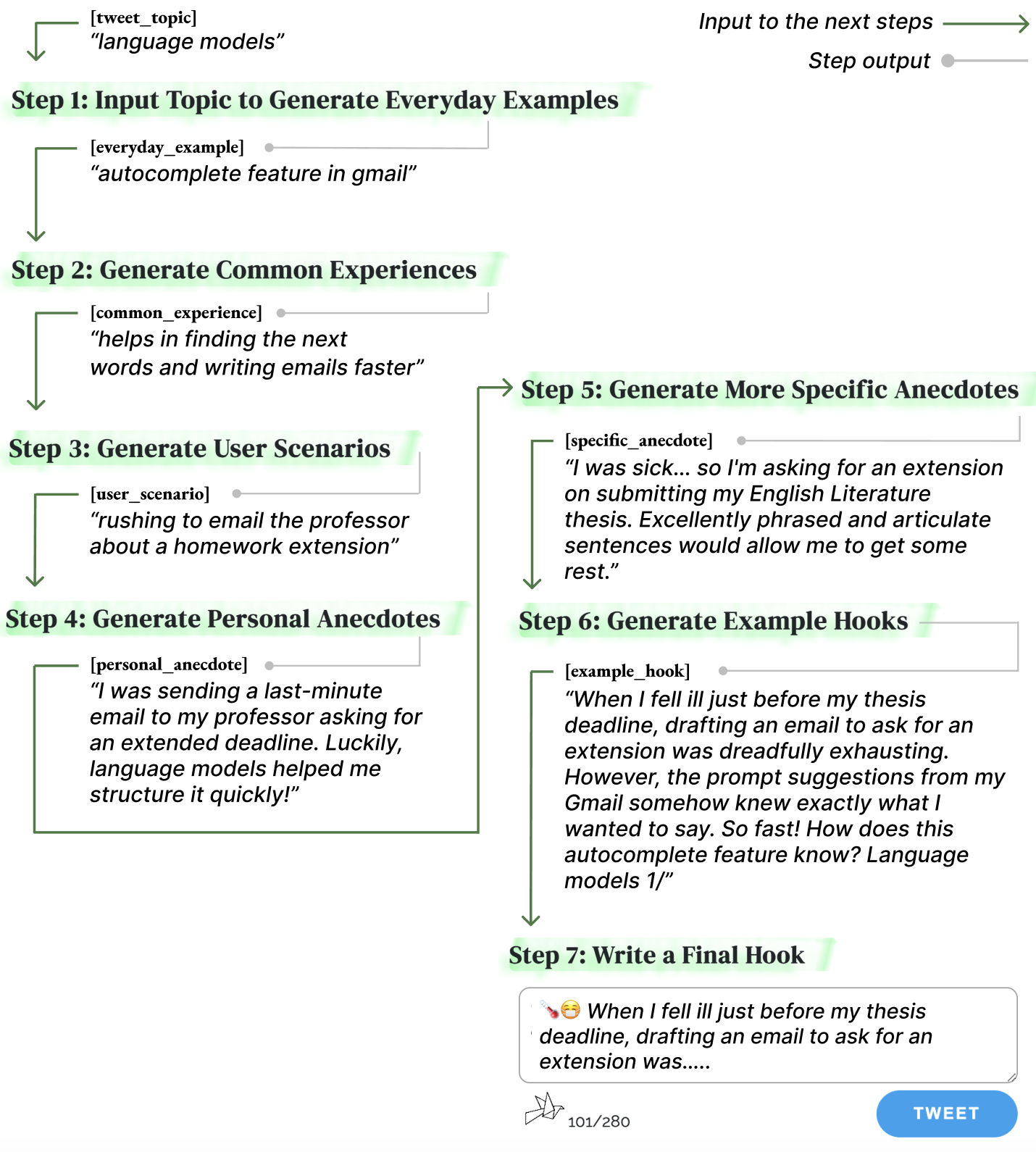}
\Description[An illustration of the existing AI workflow for Tweetorial hook writing]{An illustration of the scaffolding workflow by making science more and more contextualized: First, the user inputs their topic (e.g., language models), and the system generates 5 everyday examples and the user selects or edits one (e.g., “autocomplete feature in gmail”). Based on the topic and example, the system generates 5 common experiences and the user picks one (e.g., “helps in finding the next words and writing emails faster”). Next, the system generates 5 user scenarios where people might have this experience (e.g., “rushing to email the professor about a homework extension”). The user can again select or edit one. Next, the system generates a personal anecdote (e.g., “I was sending a last-minute email to my professor asking for extended deadlines...”). Then, based on it, it continued to contextualize and generate a more specific anecdote about the user scenario, which includes more details to make users feel relatable and personal (e.g., “I was sick... so I'm asking for an extension on submitting my English Literature thesis..”). Lastly, the system consolidates details from all the previous steps to generate example hooks, allowing users to either use them directly or start from them in the final step to write a hook.}
\caption{An illustration of the existing AI workflow for writing Tweetorial hooks, following 
\cite{Tweetorial_ICCC}. For each step inside the workflow, users
can regenerate, modify, or accept the workflow suggestions before going to the next steps.}
\label{fig:workflow}
 \end{center}
 \vspace{-30px}
\end{figure}

A 2023 study \cite{Tweetorial_ICCC} presents a human-AI workflow that scaffolds the process of writing a Tweetorial hook. As shown in Figure \ref{fig:workflow}, the workflow has seven steps and uses AI chains to help with the generations among them. The user first inputs their topic (e.g., ``\textit{language models}"), and the system generates five everyday examples, and the user selects or edits one (e.g., “\textit{autocomplete feature in gmail}”). Based on the topic and example, the system generates five common experiences, and the user picks one (e.g., “\textit{helps in finding the next words and writing emails faster}”). Next, the system generates five user scenarios where people might have this experience (e.g., “\textit{rushing to email the professor about a homework extension}”). The user can again select or edit one. Next, the system generates a personal anecdote (e.g., “\textit{I was sending a last-minute email to my professor asking for extended deadlines...}”). Then, based on it, it continued to contextualize and generate a more specific anecdote about the user scenario, which includes more details to make users feel relatable and personal (e.g., “\textit{I was sick... so I'm asking for an extension on submitting my English Literature thesis..}”). Lastly, the system consolidates details from all the previous steps to generate example hooks, allowing users to either use them directly or start from them in the final step to write a hook.

This workflow was then validated by a user study (n=10) comparing the hook-writing experience with and without the system \cite{Tweetorial_ICCC}. The workflow was shown to significantly reduce mental demand, effort, and frustration while improving performance. However, this study was conducted in a single 60-minute session, while the novelty of the workflow might have influenced the users' perception. Therefore, we conducted a study over three weeks to understand users' longitudinal interaction with the workflow.

\onecolumn
\section{System}

\begin{figure*}[!h]
 \vspace{-10px}
\includegraphics[width=.83\linewidth]{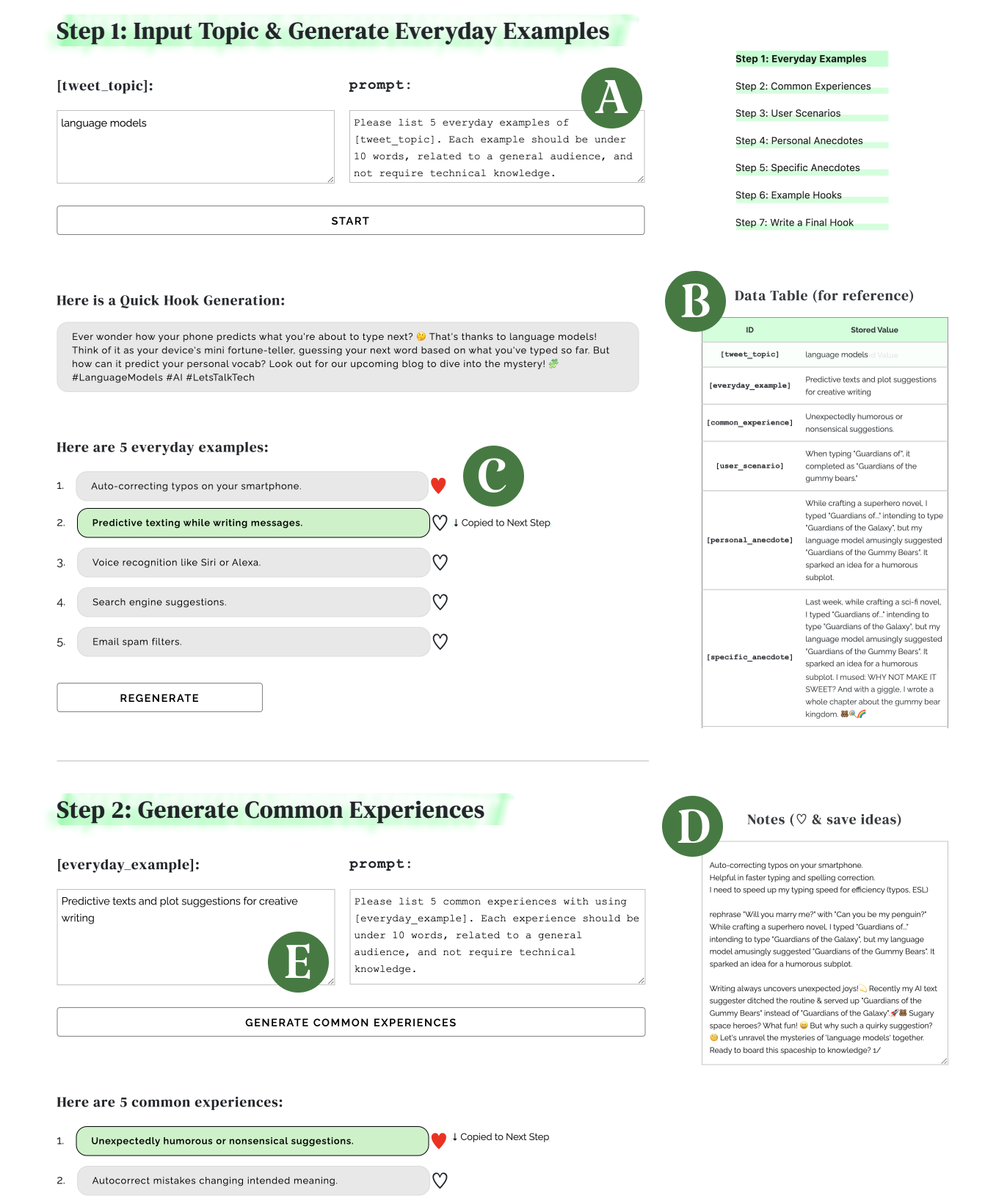}
\Description[A system diagram]{A system diagram that illustrates the first two steps of the study system along with user interactions. In this example, the user creates a Tweetorial about "language models" using the system. They input their topic and review the generated prompt. If satisfied, they press "START" to initiate the generation process. The tweet topic is automatically stored in Data Table. Next, the user examines the Quick Hook Generation but decides not to proceed due to a lack of relatability and interest in the "personal vocab" focus. Moving on to the 5 everyday examples, the user finds the autocorrecting typos relevant and bookmarks one. This bookmarked generation is copied to the Notes area for future use. The user explores the "Predictive Testing While Writing Messages" option, finding it interesting but wanting to add their creative writing experiences. They press the generated result, which is copied to input bot [everyday\_example] in Step 2. The user makes edits to the result and continues with the following steps.}
\label{fig:SystemImage}
\caption{A screenshot illustrates the first two steps of the study system, showcasing user interactions. In this example, the user tries to create a Tweetorial hook about "\textit{language models}" using the system. First, they input their topic and review the generation prompt (A). If satisfied, they press "START" to initiate the generation process. The tweet topic is automatically stored in Data Table (B). Next, the user examines the Quick Hook Generation but decides not to proceed due to a lack of relatability and interest in the "\textit{...personal vocab...}" focus. Then, moving on to the 5 everyday examples (C), the user finds the example of "\textit{Auto-correcting typos}" relevant and bookmarks it. This bookmarked generation is copied to the Notes area (D) for future use. However, after reviewing all five examples generated, the user chooses to explore the "\textit{Predictive testing while writing messages}" example further, since they find it interesting and want to integrate their personal experiences about creative writing into it. Thus, they press the generated result, which is copied to input box [everyday\_example] in Step 2 (E). The user makes edits to the result and continues with the following steps. Note: A full system walkthrough can be found in the Appendix \ref{app}.}
 \label{fig:SystemImage}

 \vspace{-30pt}
\end{figure*}


\onecolumn
\newpage
\begin{multicols}{2}
\subsection{System Design} 

As shown in Figure \ref{fig:SystemImage}, we followed the existing workflow in section 3.3 and designed a web application as a design probe~\cite{technologyprobe} to understand users’ long-term experience with AI workflows. Specifically, we want to understand (i) whether users want to continue using the full system after repeated usage - whether the novelty wears off, and (ii) whether users will take advantage of the potential of customization in AI workflows.

To address (i), we introduced ``Quick Hook Generation'' so users see an example hook based on their input topic in Step 1. If users find that the backend model can already fulfill their needs or if the novelty effects of the workflow wear off, users can proceed directly without going through the full system. For (ii), to facilitate user customization throughout usages, we display the system's prompts with variable placeholders, such as [tweet\_topic] or [everyday\_example], thus empowering users to understand, edit, and customize their system, and promoting user control and autonomy. Starting in the third week (the eighth session), users can choose to feed their previously written hooks back into the system as training examples.
We believe these minor adjustments retain the benefits of the original workflow while augmenting the user experience for longitudinal interactions. 



We developed an interactive web application prototype utilizing Python, JavaScript, Flask, and OpenAI's GPT-4 API (version-date: Aug-14-2023 to Sep-3-2023). 
We logged all user interactions and system generations.

\section{Study}

To better understand novelty effects, familiarization and customization in generative AI workflow, this study investigates the following research questions (RQs): 
\begin{itemize}
    \setlength\itemsep{3px}
    \item \textbf{RQ 1: Familiarization}.\\ How long is the familiarization phase? How do the usefulness and performance of the system change before and after familiarization? 
    \setlength\itemsep{3px}
    \item \textbf{RQ 2: Prompt Editing}.\\How and why do users customize prompts?
    \item \textbf{RQ 3: Mental Model and Ownership}.\\ How do users' mental models, sense of ownership, and involvement with the system evolve over time?
\end{itemize}

\subsection{Participants}

We recruited PhD-level researchers in computer science (CS) who were interested in communicating their research to the public on social media. Because we wanted domain experts, we targeted senior PhD students, but anyone with at least two years of research expertise was deemed expert enough. We focused on CS as a discipline because the original workflow was shown to work for CS topics. Other fields like microbiology or chemistry were not as easy to motivate with an everyday experience. Other workflows could be designed for those fields, but to test the long-term effects of a workflow, we focused on a single domain where it has already been shown to work.

\subsubsection*{\textbf{Recruitment}} We advertised to PhD students in a CS department through department-wide emails and Slack communications. The advert described that the study would last three weeks, would require about six hours of time in total, and that participants would be compensated each week - for a total of \$150 if they completed all three weeks. All recruitment materials and processes were approved by the university IRB.

\subsubsection*{\textbf{Demographics}} As listed in Table \ref{tab:demographics}, we had 12 PhD-level students in CS to complete the three-week study. Originally, we recruited 13 participants, but P5 dropped out after day one due to personal circumstances. Out of the final 12 participants, the average age was 27.5, with six men and six women. On average, they had five years of research experience. Their expertise spans various CS research areas, including theory, systems, security, machine learning, natural language processing, and human-computer interaction. They have varying levels of experience with large language models: six out of 12 users use it weekly, four use it monthly, one uses it daily, and one has never used it. All users had frequent science communication experience, mostly at conferences. Only five participants mentioned that they tweeted about scientific topics or papers.

\end{multicols}

{
\vspace{-10px}
\begin{table}[!b]
\begin{longtable}{@{}ccccccc@{}}
\caption{Overview of participant demographics and research background} 
\Description[An overview table of participant demographics and research background]{An overview table of participant demographics and research background}
\label{tab:demographics}\\
\toprule
\textbf{Participant ID} & \textbf{Gender} & \textbf{Age} & \textbf{Research Area} & \textbf{PhD Year} & \textbf{Research Experience} & \textbf{SciComm Frequency} \\
\midrule
\endfirsthead
\endhead
\textbf{P1}  & Male    & 39  & Security and Reliability of Systems & 5th Year & 5 Years & Weekly \\
\textbf{P2}  & Male    & 25  & Theoretical Computer Science         & 4th Year & 6 Years & Weekly \\
\textbf{P3}  & Male    & 26  & Software Systems                     & 2nd Year & 4 Years & Monthly \\
\textbf{P4}  & Female  & 24  & Natural Language Processing          & 1st Year & 4 Years & Weekly \\
\textbf{P5 (Dropped)} & - & - & -                            & - & - & - \\
\textbf{P6}  & Female  & 28  & Speech Processing                    & 2nd Year & 5 Years & Weekly \\
\textbf{P7}  & Male    & 26  & System for Machine Learning          & 3rd Year & 5 Years & Daily \\
\textbf{P8}  & Female  & 24  & Natural Language Processing          & 2nd Year & 2 Years & Daily \\
\textbf{P9}  & Male    & 23  & Natural Language Processing          & 1st Year & 5 Years & Weekly \\
\textbf{P10} & Female  & 26  & Natural Language Processing          & 4th Year & 7 Years & Daily \\
\textbf{P11} & Female  & 34  & Human-Computer Interaction           & N/A      & 3 Years & Weekly \\
\textbf{P12} & Male    & 24  & Virtual Reality                      & 1st Year & 2 Years & Monthly \\
\textbf{P13} & Female  & 31  & Machine Learning Theory              & 4th Year & 10 Years & Daily \\
\bottomrule
\end{longtable}
\vspace{-25px}
\twocolumn
\end{table}

}

\newpage
\clearpage

\noindent 
 
\subsection{Procedure}
\begin{figure}[!h]
    \centering
    \includegraphics[width=1\linewidth]{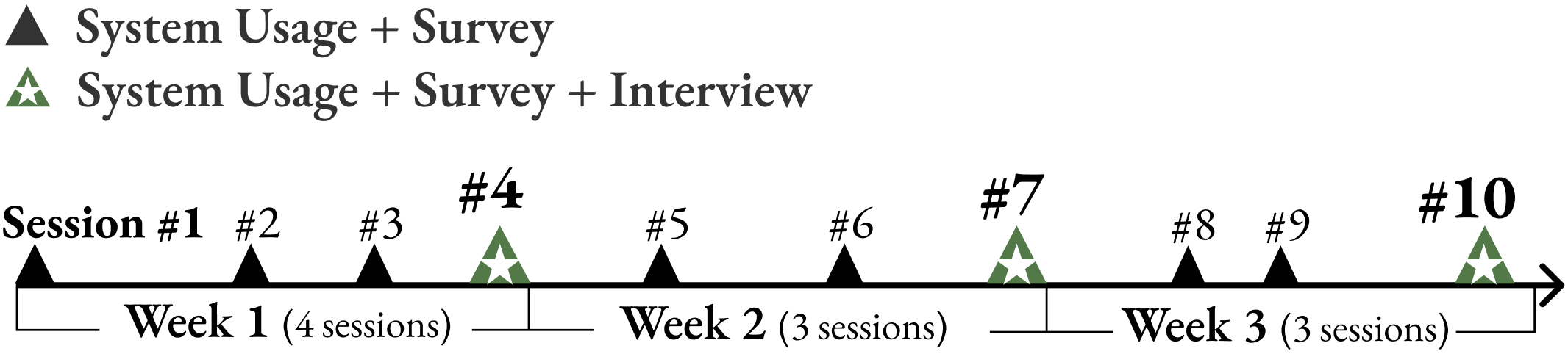}
    \Description[An illustration of the study activities timeline]{An illustration of the study activities timeline over three weeks. In the first session, users are introduced to the system and study and write a hook with the system with the study administrator present to observe the process and answer questions about the task, the system, and the study. In the remaining nine sessions, the following pattern is repeated three times: two solo system usages (without the administrator present), then a system usage with the administrator present to observe and conduct a semi-structured interview.}
    \caption{Timeline of study activities over three weeks. Each week, users need to use the system three times to write Tweetorial hooks, followed by a semi-structured interview conducted at the week's end.}
    \label{fig:studyprocedure}
    \vspace{-5px}
\end{figure}

As shown in Figure \ref{fig:studyprocedure}, all 12 participants wrote 10 hooks in 10 separate sessions over the course of three weeks. Each session lasted around 20 minutes, followed by a 10-minute survey. In the first session, users are introduced to the system and study and write a hook with the system with the study administrator present to observe the process and answer questions about the task, the system, and the study. In the remaining nine sessions, the following pattern is repeated three times: two solo system usages (without the administrator present), then a system usage with the administrator present to observe and conduct a 30-minute semi-structured interview. 

\subsection{Data Collection and Analysis}
\subsubsection*{\textbf{System Log}} We logged all 120 system usage sessions to record all interaction details, including the user's inputs and prompts for each step, the system's generation results, the system outputs they bookmarked and proceeded with, and the timestamps associated with each activity. During each system usage (either personal use or with the study administrator), we logged every interaction on the Python server backend in a separate .log file. After the user concluded the usage session, a JSON file with the same information was automatically downloaded to the user's laptop, allowing users to review their experiences if needed and ensuring transparency in data saving. Then, they need to submit the file for the session survey to verify their usage.

To analyze the log data, the first author used Python to analyze the user's session interactions, including extracting the duration spent on the system and each step, a list of steps taken, skipped, or revisited by users, and the input data and prompts utilized for each step's generation. Then, to understand how users make prompt edits, the first author used a collaborative affinity diagramming tool to group users' new prompts into subgroups and groups based on their editing objectives with an inductive approach. The final (sub)groups and their examples were iteratively refined and discussed, with feedback sought from the third author.

\subsubsection*{\textbf{Survey}} We received 118 surveys for the 120 system usage sessions\footnote{
Each user was supposed to submit 10 surveys, resulting in a total of 120 surveys. However, P10 and P13 missed their survey submission for their first usage sessions. The study administrator later followed up with them about their experiences during week one interviews, and both users shared they didn't notice any significant difference between their first and second usage.} to understand users' task experience, perceptions of the system's usefulness, and performance. After each system usage session, users were asked to complete a Google Form survey. After submitting their participant ID and their session JSON files, users needed to fill out three sections:
Section A focused on the hook writing task experience alone and included questions adapted from the NASA Task Load Index (NASA TLX) \cite{nasatlx}.
Section B primarily focused on their system experience, probing participants' perceptions regarding the system’s usefulness, mental model, and expectations for future uses. 
Lastly, Section C had users self-assess the quality and effectiveness of their hook, followed up with a question on the level of ownership they felt over it.

To analyze the survey data, the first author conducted exploratory data analysis using Python to see the relationship between the variables and uncover relationships among session experiences, system performance, user perceptions, and end output evaluations over time and novelty. Then, the first author conducted statistical testing to test statistical significance.

\subsubsection*{\textbf{Interview}} We conducted 36 interviews with users to understand their degree of familiarity with the system and other experiences and thoughts. Before each interview, users first engaged in a system usage session with the study administrator, during which they were encouraged to think aloud, express their feelings, explain their rationales behind specific choices, and detail their overall experience with the system. Following this, we conducted semi-structured interviews to understand their experiences from the past week, covering any noticeable shifts in their experiences, as well as their expectations for the upcoming week regarding workflow and experiences. We audio-recorded and transcribed the interview sessions using Zoom's auto-transcription feature after receiving the user’s consent each time.

To analyze the interview data, the first author performed qualitative coding on the interview transcripts, which were then verified by the second and third authors. The qualitative coding follows the inductive thematic approach~\cite{fereday2006demonstrating} to discern patterns and shifts in user experiences, building an affinity diagram based on the codes. The first author then iterated on the codes and (sub)themes to answer research questions, reporting to the other two authors more than five times for refinement.

\section{Results}

\begin{table*}[b]
\vspace{5px}
{
    \centering
    \Description[A table showing the number of sessions until the familiarization phase concludes]{A table showing the number of sessions until the familiarization phase concludes}
    \caption{Number of sessions until the familiarization phase concludes.}
    \label{tab:learningPhase}
    \begin{tabular}{cc}
         \toprule
        Participant ID & Familiarization concluded by \# sessions \\
        \midrule
        P1  & 5 \\
        P2  & Didn't conclude familiarization\\
        P3  & 3 \\
        P4  & 6 \\
        P6  & 4 \\
        P7  & 4 \\
        P8  & 4 \\
        P9  & 4 \\
        P10 & 5 \\
        P11 & 4 \\
        P12 & 5 \\
        P13 & 3 \\
        \midrule
        \textbf{Average} & \textbf{4.27} \\
        Median & 4 \\
        \bottomrule
    \end{tabular}
    }
\end{table*}

\subsubsection*{\textbf{Initial User Experience}} 
Based on the initial survey responses and insights gathered from week one interviews, all 12 users had positive experiences with the system during their first system usage session. 
Specifically, the users already found the Tweetorial hook writing task challenging and found the system useful in supporting the task. Notably, the ratings for the system were already high: the average rating for the writing task experience was 5.72 out of 7, while the perceived system usefulness scored 5.82 out of 7. Nine users also mentioned the system helped them better express themselves during the first week interview. These comparatively high starting ratings aligned well with findings from earlier research on workflow and the novelty effect, as users might find such automated systems new and "magical" only in the early phase. Thus, the following section addresses the following questions: Were these positive reactions primarily driven by a novelty effect? And more importantly, would users' perceptions of the system's usefulness persist over time?

\section*{RQ 1: Familiarization}
How long is the familiarization phase (RQ 1.1)? How do the usefulness and performance of the system change before and after familiarization (RQ 1.2 and RQ 1.3)?

\vspace{3px}

\subsection*{RQ 1.1 How long is the familiarization phase?}
\vspace{3px}
\noindent\fbox{
    \parbox{.95\linewidth}{
\textit{Summary: On average, the familiarization ends after 4.27 sessions. Before familiarization, users were exploring to gain a better understanding of the system. After familiarization, users understood the workflow, were able to anticipate the workflow outputs, and finalized their own versions of the workflow. These changes influenced users' perceptions of the system's usefulness and performance.}
}
}
\vspace{12px}

To understand when users became familiar with the system, we asked them to reflect on their experience and the degree to which they are familiar with the system during our week two and week three interviews (after sessions 7 and 10). 
During the two interviews, we first asked, "Do you feel you fully understand the workflow now?" Then, if the user shared that they had already mastered the system, we followed up with the question, “At which session do you think you fully understood the systems?” 
On average, users identified their familiarization phase ending at 4.27 sessions (median: 4 sessions). This equates to 1.42 hours of system usage, and it was achieved in the first half of week 2 of the study (Table \ref{tab:learningPhase} shows the specific session number for each user needed to end familiarization).

We asked the users to characterize their experiences and interactions before familiarization. All 12 users shared that they were exploring and attempting to understand the system. Four users (P1, P3, P7, and P11) described that they were "going with the workflow” rather than trying to edit or critique it. Similarly, three users (P4, P9, and P13) chose from the options provided with little editing. 
P8 expressed it as follows:
\vspace{2px}
\begin{quote}
\begin{quote}
\textit{"I feel like [the familiarization] ends early this week, [so after] the fourth usage. I just remembered what each step is and... understood it in the way, not from the definition of the step, but also from the generation results of the step. So that kind of helped me to climb through the learning curve..."
}
\begin{flushright}--- P8, Week 2 Interview\end{flushright}
\end{quote}
\end{quote}

After familiarization, users expressed three key changes in their experiences. 
First, users mentioned they understood each step more concretely and how the steps build towards a final hook. 
Second, users were able to anticipate system generations. They could more quickly decide which steps to skip and which steps to spend more time on based on their importance to the outcome and the abilities of the LLM.
Third, ten users felt they "finalized" their own version of the workflow. They identified their preferred ways of using the system and the prompt edits and strategies that work best for their respective topics.  P3 expressed it as follows:

\begin{quote}
\begin{quote}

\textit{“[Now I] have my own fixed routine to generate hook… For me, the workflow starts to converge… the output is satisfying for me, so I kinda stop exploring a new way to improve the result. I have a fixed workflow… so my workflow stops changing.”}
\begin{flushright}--- P3, Week 3 Interview\end{flushright}

\end{quote}
\end{quote}

One of the 12 users, P2, did not reach familiarization within the ten sessions of the study. When reflecting on why the system didn't work, they said that their topics were in an area of theoretical computer science, which lacked obvious everyday use cases that the system was designed to support. Then, they thought analogies were a more natural fit for motivating his theoretical topics.  
Thus, they tried adapting the workflows to generate analogies instead of relying on the use case mechanism and thought that was promising. We return to this challenge in the Discussion section 7.2. 
\vspace{2px}

\begin{figure*}[t]
\centering
\includegraphics[width=.93\linewidth]{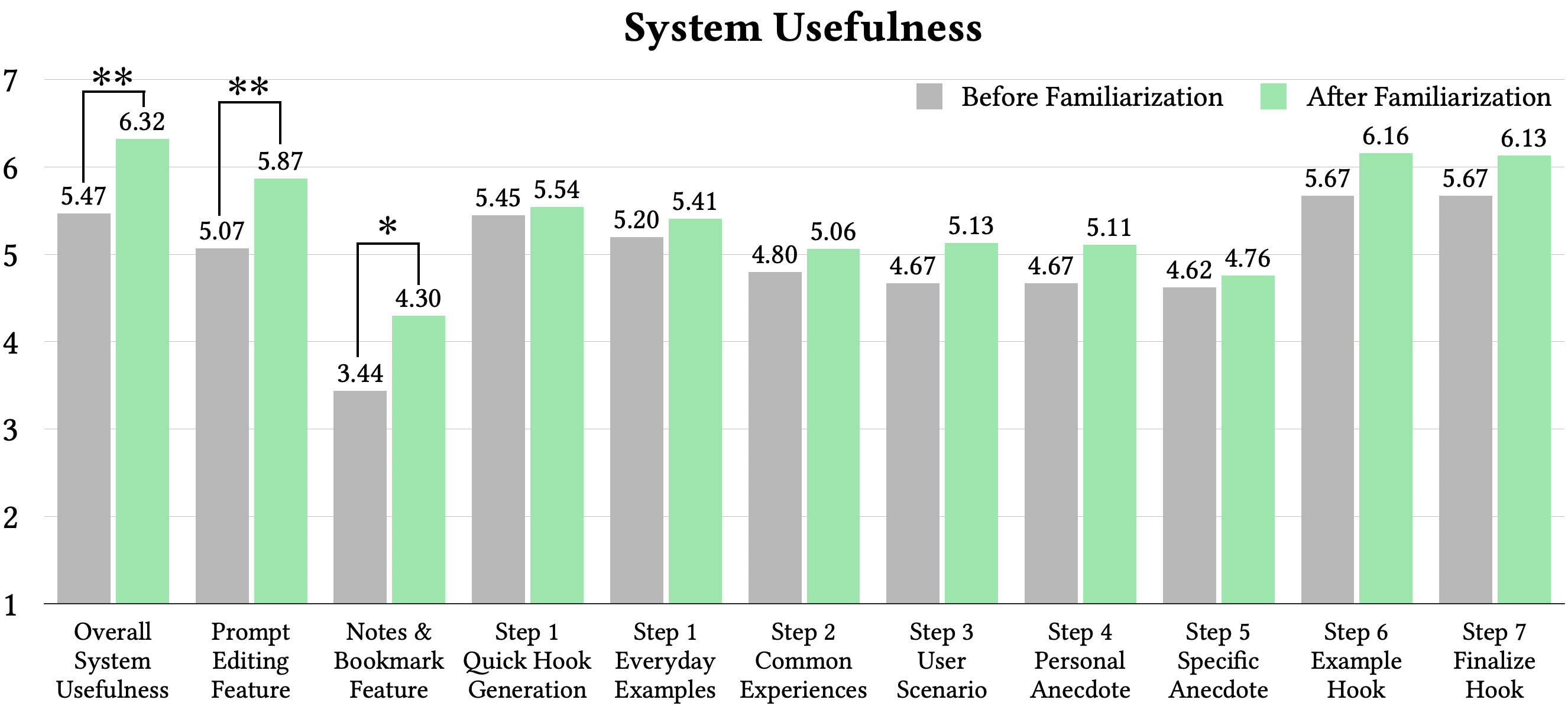}
\Description[A bar chart showing system usefulness changes before and after familiarization]{A bar chart showing system usefulness changes before and after familiarization. The bars inside the graphs are Overall System Usefulness, Prompt Editing, Note Section, Step1 Quick Hook, Step1 Everyday Example, Step2, Step3, Step4, Step5, Step6, and Step7. Then,
the average scores from 12 users indicate statistical significance before and after familiarization only within Overall System Usefulness, Prompt Editing, and the Note Section. For the Overall System Usefulness and Prompt Editing, the p-value is smaller than 0.005 while the p-value for the Note Section is smaller than 0.05.}
\caption{System Usefulness (average scores across 12 users before and after familiarization, ** denotes statistical significance at the p-value < 0.005 level, * denotes statistical significance at the p-value < 0.05 level)}
\label{fig:usefulness}
\vspace{10px}
\end{figure*}

\subsection*{RQ 1.2 System usefulness before and after familiarization\footnote{\label{note1}Based on each user's familiarization phase cutoffs, we categorized 120 user sessions into two phases: \texttt{'Before Familiarization'} with 57 sessions and \texttt{'After Familiarization'} with 63 sessions for our log analysis and survey results. We put all ten sessions of P2 under the `Before Familiarization' tag, as P2 indicated that they did not reach familiarization within the entire study.}}

\vspace{6px}
\noindent\fbox{
    \parbox{.96\linewidth}{
    \textit{Summary: The overall system usefulness improved by 12.1\% after familiarization. The only features that show an increase in usefulness are the prompt editing feature and the notes \& bookmark feature, which serves as a support for prompt editing. In contrast, the usefulness of all the individual steps within the workflow (Step 1 to 7) remained unchanged.
    }
}}

\vspace{6px}

Based on the session survey results shown in Figure \ref{fig:usefulness}, users reported a 12.1\% (0.85/7) improvement in overall system usefulness before and after familiarization (5.47/7 to 6.32/7, p-value < 0.005). 
This indicates that there is no novelty effect --- over time, workflow utility increases rather than decreases.

When analyzing where the benefits come from, we analyzed the survey on the usefulness of specific system features. On average, there is no change in the perceived usefulness of the individual system steps (Step 1, Step 2,... , Step 7). By week 3, most users still felt every step was  “\textit{useful and meaningful} (P1),” mostly by providing an easy "\textit{direction to start} (P6)". At the conclusion of the study, P12 continues to find all the steps and the system useful:

\vspace{8px}
\begin{quote}
    \begin{quote}
        \textit{"Even in Steps 1, 2, 3, I'm finding that I need to regenerate a lot to get what I want. It's still much easier than coming up with everything on my own. [...] I don't think there's a single step that I don't find useful. Even Step 7 is useful because it gives me like a word counter for tweets."}
        \begin{flushright}
            --- P12, Week 3 Interview
        \end{flushright} 
    \end{quote}
\end{quote}
\vspace{6px}

Instead, the system usefulness data showed only two features with statistically higher perceived usefulness: prompt editing with an 11.4\% improvement (5.07/7 to 5.87/7, p-value < 0.005) and notes \& bookmarks with a 12.3\% improvement (3.44/7 to 4.3/7, p-value < 0.05). 
Both of these features are essential for customizing the workflow. Through prompt editing, users can tailor the system to their style and topic. The bookmarks feature allows people to "heart" their favorite generation, enabling them to explore multiple generations before choosing the best one. As users generate customized prompts and generate more, bookmarks become increasingly important. Thus, it seems that two customization-related features are mostly responsible for the increase in overall system usefulness. For example, P13  noted this association in their week three interview:
\vspace{3px}
\begin{quote}
    \begin{quote}
 \textit{"I now better know how to deal with the prompt [editing] and the [bookmark \& save feature] box on the left side… It was useful... After learning how I can edit the prompt, yes, [the system gets] more useful."}
     \hspace{35px}  --- P13, Week 3 Interview
    \end{quote}
\end{quote}
\vspace{8px}

\subsection*{RQ 1.3 Performance before and after familiarization\footnotemark[2]}
\vspace{2px}

\begin{figure*}[t]
\centering
\begin{minipage}{.544\linewidth}
  \centering
  \includegraphics[width=\linewidth]{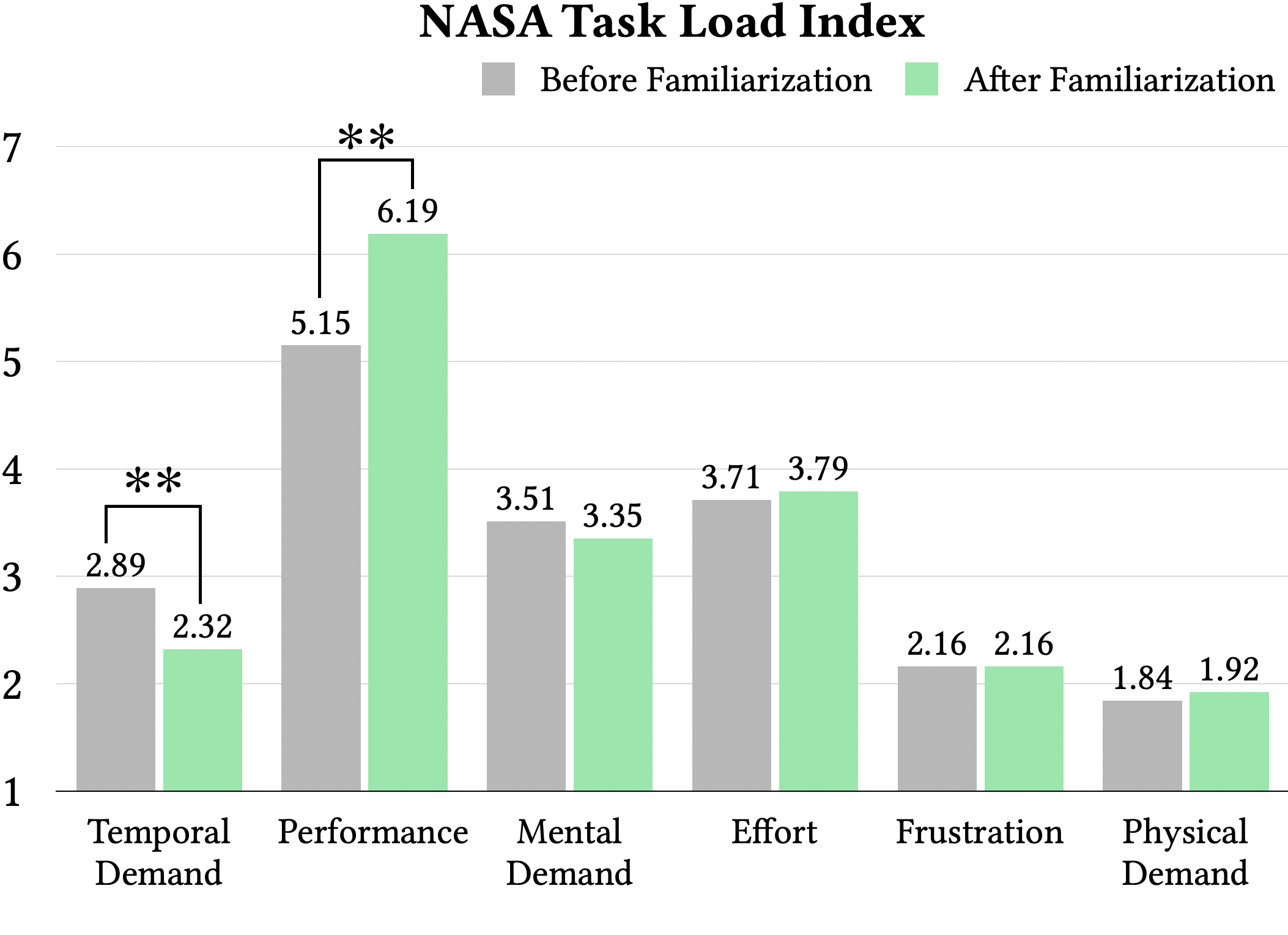}
  \Description[A bar chart showing users' NASA Task Load Index before and after familiarization]{A bar chart showing users' NASA Task Load Index before and after familiarization. The bars inside the graphs are Temporal Demand, Performance, Mental Demand, Effort, Frustration, and Physical Demand. Then, the average scores from 12 users indicate statistical significance before and after familiarization only within Temporal Demand and Performance, for which p-values are both smaller than 0.005.}
  \caption{NASA Task Load Index (average scores across 12 users before and after familiarization, ** denotes statistical significance at the p-value < 0.005 level)}
  \label{fig:nasa}
\end{minipage}%
\hspace{0.04\textwidth}
\begin{minipage}{0.38\linewidth}
  \vspace{11.5pt}
  \includegraphics[width=\linewidth]{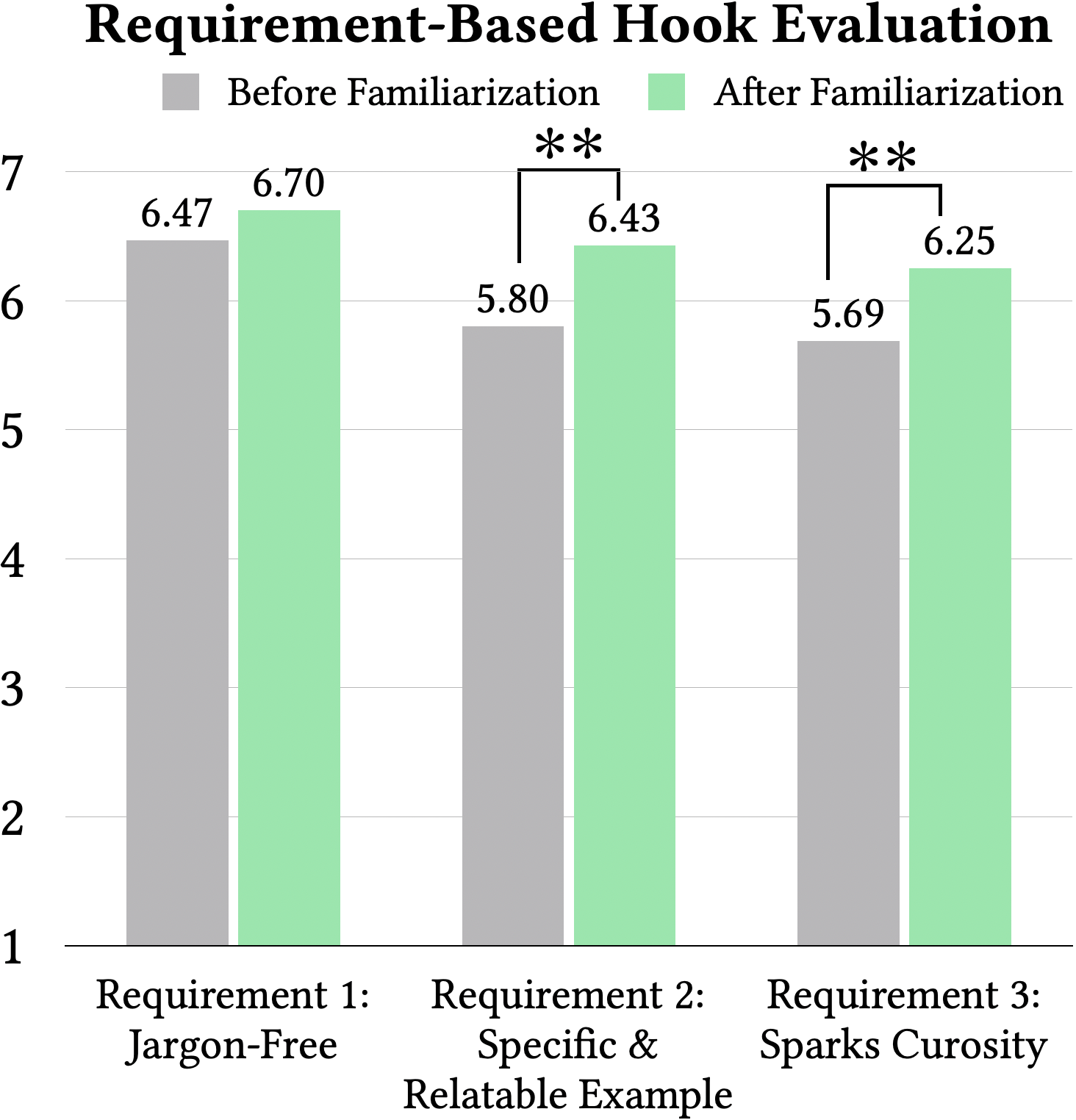}
  \Description[A bar chart showing users' hook evaluation scores based on three requirements]{A bar chart showing users' hook evaluation scores based on three requirements before and after familiarization. The bars inside the graphs are Requirement 1 No Jargon, Reuqirement 2 Relatable and Specific Example, and Requirement 3 Sparks Curiosity. Then, the average scores from 12 users indicate statistical significance before and after familiarization only within Requirement 2 and Requirement 3, for which p-values are both smaller than 0.005.}
  \caption{Hook evaluation based on three requirements (average scores across 12 users before and after familiarization, ** denotes statistical significance at the p-value < 0.005 level)}
  \label{fig:hookeval}
\end{minipage}
\hspace{0.02\textwidth}
\vspace{-3px}
\end{figure*}

\noindent\fbox{
    \parbox{.96\linewidth}{
    \textit{Summary: Users reported a 14.9\% improvement in their task performance and a 7.1\% improvement in time efficiency. From their self-assessments, users believe they produced better hooks, with more specific and relatable examples (Requirement 2) and better at sparking curiosity (Requirement 3).
}
}}
\vspace{3px}

From the NASA TLX scores reported from the surveys, users noticed a 14.9\% improvement in performance (5.15/7 to 6.19, p-value < 0.005) and a 7.1\% improvement in time efficiency for completing the task (2.89/7 to 2.32/7, p-value < 0.005) after familiarization. However, as shown in Figure \ref{fig:nasa}, the task remained similarly mentally demanding, effort-demanding, and frustration-inducing over the two stages, with changes of less than 0.2. Paired t-tests further confirmed that these marginal differences were not statistically significant. 
This is consistent with RQ 1.1 that people explore less after familiarization, thus saving them time but not necessarily further reducing mental load. For example, P6 shared in the week three that they noticed an improvement in temporal demand:

\vspace{2px}
\begin{quote}
    \begin{quote}
\textit{"I used the system much faster this week... For the first week, I tried to understand what [each step] means... for the last week, I already knew what it means... so I chose the [generation result] option quicker than at the beginning."} 
\begin{flushright}
    --- P6, Week 3 Interview
\end{flushright}
\end{quote}
    \end{quote}
\vspace{6px}

Based on users' self-assessments of their final hooks, they first reported producing better hooks after the familiarization phase. In Figure \ref{fig:hookeval}, users shared a 9\% improvement for \textit{Requirement 2: Specific and Relatable Example(s)} (from 5.80/7 to 6.43/7, p-value < 0.005) and an 8\% improvement for \textit{Requirement 3: Sparks Curiosity }(from 5.69/7 to 6.25/7, p-value < 0.005).
Also, users noted a 17.3\% reduction in the frequency of hooks containing factual inaccuracies, biases, or potential risks. Before familiarization, 68.4\% of hooks were labeled as accurate and unbiased (39 out of the 57 hooks). After familiarization, this increased to 85.7\% (54 out of 63 hooks).


This again argues that familiarization with the AI workflow was not just a novelty effect but provided genuine value to users.

\section*{RQ 2: Prompt Editing}
\subsection*{How and why do users customize prompts?}
\noindent\fbox{
    \parbox{.96\linewidth}{\textit{Summary: Users customized 12.4\% of the available prompts (89 of 720). These edits addressed three main problems with the AI workflow: going off-topic, undesired narrative and style, and inefficient workflow. In week 3, we added users' own hooks into the prompt as training examples, but there was no significant increase in the utility of the system. Nine users favored this new feature, while three reported no benefits.}
}}

\subsection*{Prompt Editing Themes \& Strategies}

Across 120 sessions, all users made a collective total of 89 prompt edits. One author coded the 89 edited prompts using the inductive thematic analysis and reported to another author for feedback and iterations on codes. Table \ref{tab:strategies} shows the six prompt editing strategies that the users used to solve three workflow problems:

\subsubsection*{\textbf{Problem \#1: Off-Topic}}
The system occasionally failed to maintain a clear topic definition or emphasis over generations. To rectify this, users either reiterated the topic for relevance or added definitions and more details:

\vspace{5px}
1a. Reiterate Topic (22 edits):
When the system veered off the technical topic during a lengthy generation sequence, users reinforced the theme. They either reiterated their Tweetorial topic or inserted the reference ID, like [tweet\_topic], into the prompt to keep generations relevant.

\vspace{5px}
1a. Clarify Definition (5 edits):
Occasionally, users perceived the system as lacking depth or contemporary knowledge about specific topics. To address this, they either embedded a direct definition or sourced definitions from online platforms like Wikipedia.

\clearpage
\newpage

{
\clearpage
\onecolumn

\begin{multicols}{2}
\subsubsection*{\textbf{Problem \#2: Undesired Narrative and Style}}

The system's outputs sometimes didn't align with users' desired narrative or stylistic preferences. To personalize and guide the generations, users either add preferred examples as training to guide generations or provide longer instructions:

\vspace{5px}

2a. Guide Narrative (25 edits):
The system sometimes missed users' preferred narratives or specific nuances they wanted to highlight. Users then reinforced their desired narrative direction by adding specifics or personal experiences.

\vspace{5px}
2b. Guide Style (18 edits):
To better mirror their unique writing styles, users made adjustments. Whether they wanted casual language, content tailored for a teenage audience, or the inclusion of emojis, users fine-tuned the prompt to achieve the desired output.

\subsubsection*{\textbf{Problem \#3: Inefficient Workflow}}
The system sometimes failed to generate users' anticipated outcomes efficiently after familiarization. To utilize preferred, shorter version of the workflow or explore more options at once, they either sidestepped some middle steps or requested the system to diversify generations:

\vspace{5px}
3a. Increase Variations (14 edits):
Users observed that the original system prompt often generated a restricted number of examples. To expand their options, they modified the prompt to produce between 10 and 20 example outputs for broader brainstorming.

\vspace{5px}
3b. Skip Steps (5 edits):
When users identified valuable ideas early on, they sometimes wished to advance without adhering to every step. To streamline their process, they altered the system's reference ID to reflect the results of the previous step.

\end{multicols}

\vspace{-20px}
\renewcommand{\arraystretch}{1.8}
\begin{table*}[h]
\Description[A table showing the 6 prompt editing strategies and their examples.]{A table showing the 6 prompt editing strategies and their examples.}

\begin{longtable}{m{0.1\linewidth}m{0.15\linewidth}m{0.68\linewidth}}
\label{tab:strategies}\\
\caption{Overview of the 6 prompt editing strategies and their examples.}\\
\toprule
\multicolumn{1}{p{0.09\linewidth}}{\textbf{Problems}} &
\multicolumn{1}{p{0.15\linewidth}}{\textbf{Editing Strategies}} &
\multicolumn{1}{p{0.68\linewidth}}{\textbf{Example} \newline (Note: \textcolor{purple}{\textbf{Highlights}} indicate user additions or deletions (\st{strikethrough}) to the original prompt)\vspace{2pt}} \\
\midrule
\multirow{2}{*}{\parbox{2.2cm}{\textbf{\\\\Problem 1:\\Off-Topic}}} & \parbox{2cm}{\centering \vspace{-10px}Reiterate Topic\\(22 edits)}               & Please list 5 user scenarios someone encounters with [common\_experience] \textcolor{purple}{\textbf{and [tweet\_topic] and augmented reality}}. Each scenario should be under 15 words, from a first-person perspective, related to a general audience, and not require technical knowledge. (P12, Step 3: Generate User Scenarios)\vspace{2pt}    \\    \cline{2-3}

                           & \parbox{2cm}{\centering \vspace{-6px}Clarify Definition\\(5 edits)}               & Please list 5 everyday examples of [tweet\_topic]. \textcolor{purple}{\textbf{[tweet topic] is defined as using computational methods, often natural language processing, to analyze social science such as literature.}} Each example should be under 10 words, related to a general audience, and not require technical knowledge. (P4, Step 1: Generate Everyday Examples)\vspace{2pt}    \\ \midrule
\multirow{2}{*}{\parbox{2cm}{\textbf{Problem 2: \newline Undesired \\Narrative\\and Style}}} & \parbox{2cm}{\centering \vspace{-14px}Guide Narrative\\(25 edits)}             & Here is a personal anecdote about the topic, [tweet\_topic], for the experience of [common\_experience]: [specific\_anecdote]. Based on this anecdote, please write 3 versions of the opening tweet to introduce the concept of [tweet\_topic]. The tweet should have no jargon. It should contain a specific and relatable example. It should have a driving question to spark curiosity and make readers want to read more. \textcolor{purple}{\textbf{It should emphasize giving everyone an equal number of candies in broad categories, like chocolates, caramels, and that some candies can be in multiple categories.}} (P2, Step 6: Generate Example Hooks) \vspace{2pt}   
\\ \cline{2-3} 
                           & \parbox{2cm}{\centering \vspace{-6px} Guide Style\\(18 edits)}               & Please list 5 common experiences with using [everyday\_example]. Each experience should be under 10 words, \textcolor{purple}{\textbf{related to a teenage audience,}} and not require technical knowledge. (P4, Step 2: Generate Common Experiences) \vspace{2pt}    \\\midrule
\multirow{2}{*}{\parbox{2.2cm}{\textbf{\\ \\Problem 3:\\ Inefficient\\Workflow}}} & \parbox{2.1cm}{\centering \vspace{-10px}Increase Variation\\(14 edits)} & Please list \textcolor{purple}{\textbf{\st{5} 20}} user scenarios someone encounters with [common\_experience]. Each scenario should be under 15 words, from a first-person perspective, related to a general audience, and not require technical knowledge. (P9, Step 3: Generate User Scenarios)  \vspace{2pt}    \\ \cline{2-3} 
                           &  \parbox{2cm}{\centering \vspace{-8px} Skip Step\\(5 edits)} & Please write 3 short personal anecdotes about the topic, [tweet\_topic],\textcolor{purple}{\textbf{\st{ for the user scenario of [user\_scenario]}}}. Each anecdote should be under 30 words, from a first-person perspective, related to a general audience, and not require technical knowledge. \textcolor{purple}{(*Note: The user skipped three steps from Step \#1 to \#4, so they only used [tweet\_topic] for this step, without chaining the [user\_scenario].)} (P7, Step 4: Generate Personal Anecdotes) \vspace{2pt}       \\ \bottomrule
\end{longtable}

\end{table*}
\twocolumn
\vspace{-20px}
}

\subsection*{Augmenting Prompts with Users' Hooks}

Starting from the 8th session (day one of week three), we introduced a new feature: the Learn-My-Style option inside `Step 6: Generate Example Hooks'. 
This step allows the user to see two versions of the prompt: one is the original generate-example-hook prompt, and the other combines the original prompt with their own hook examples taken from their previous seven sessions. Users have the option to select one of the prompts to proceed, and they can also edit their own hook examples within the prompt if they wish (See the updated interface in Appendix \ref{app1}).

In theory, this Learn-My-Style prompt could enhance the system's generations by better adhering to users' writing styles. It also has the potential to provide more context about the topics or expertise areas users usually select from.

However, the session surveys revealed no statistically significant difference in users' perceived system usefulness and performance before and after we implemented the new feature. Specifically, between week two (sessions 5, 6, 7) and week three (sessions 8, 9, 10), there was no statistically significant difference in perceived usefulness for the overall system (6.03/7 - 6.06/7, p-value = 0.87) and for Step 6 (5.83/7 - 6.08/7, p-value =  0.47). Additionally, there was no statistically significant difference in performance (5.91/7 - 6.03/7, p-value = 0.40).

During the final interview, nine users expressed positive opinions on this feature, while three others showed either indifference or did not like the addition. 
All nine users who favored the new prompt compared the generated hooks with and without adding their own hook training data to understand how it works. Among them, four noted that their unique style was reflected in the generated hooks. For instance,  
P4 stated their preference for humor and darker themes, commenting that the hooks generated by the new prompt resonated well with their taste. P7 found the new prompt successfully captured their habit of ending tweets with a question and avoiding direct mention of topic names. P10 mentioned they like using personal pronouns, like "you," "she," "my," etc., to make hooks more relatable, and they saw this trait also from using the Learn-My-Style prompt. 

Among the users less enthusiastic about this feature, P6 expressed hesitancy, feeling they needed more time for familiarization. P13 preferred using the original prompt, finding its simplicity and current results satisfactory. P9 offered a unique perspective: they saw the system as a tool to explore styles beyond their own. They believed the original system could offer a broader creative spectrum than their own style while incorporating their previous generations inside the prompt might constrain creativity and limit variety:

\vspace{3px}
\begin{quote}
    \begin{quote}
\textit{"I [already] had some biases in selecting hooks... [When I used] the original hook system, it might have just like more creative [and] expressive results. But then, I also tried to use the Learn-My-Style thing,      and I was like, yeah, the creativity does get hindered."}
\hspace{52px}--- P9, Week 3 Interview

\end{quote}
    \end{quote}
\vspace{2px}

In conclusion, prompt editing becomes more useful over time. After users learn how and what they can add to prompts to enhance them, they report finding it significantly more valuable. Users were divided on whether adding their hooks into the prompt as an example was beneficial. Even for those who found it useful, it addressed surface-level aspects like style rather than deeper structural issues like constraining the generations or skipping steps. Thus, AI generation systems would likely benefit users in the long term by allowing end users to edit prompts, enabling them to make structural changes to the system.

\section*{RQ 3: Mental Models and Ownership}
\subsection*{How have users' mental models, ownership, and involvement changed over time?}
\vspace{3px}
\noindent\fbox{
    \parbox{.96\linewidth}{\textit{Summary: Users' mental models remain fairly consistent over time. However, users exhibit different levels of involvement and ownership throughout longitudinal interactions. Eight of 12 users thought that more involvement with the system led to increased feelings of ownership.}}}
\vspace{5px}
\subsection*{Mental Models}

We analyzed the survey question on how users would describe the system throughout the 120 sessions. They primarily viewed the system as a tool (90.7\% of the sessions, 107 times) and content generator (84.7\% of the sessions, 100 times), with less frequent perceptions like co-pilot (44.1\% of the sessions, 52 times), reference guide (38.1\% of the sessions, 45 times), and ghostwriter (28.8\% of the sessions, 34 times).

Before and after familiarization, there is no significant change in the user's mental model of the system (See Table \ref{tab:toolComparison}). After familiarization, there was a small increase in references to the system as a "content generator" and "collaborator", with increases of 11.70\% and 11.45\% respectively. Descriptions of the system as a "tool" decreased by 7.27\%. However, these shifts were subtle and not statistically significant. This indicates that users maintain a consistent perception and mental model of the system before and after becoming familiar with it. This suggests that when the novelty wears off, users don’t usually process new mental models for such a customizable system. 

\begin{table}[ht]
\Description[A table showing users' mental model of the system before and after familiarization.]{A table showing users' mental model of the system before and after familiarization (percentage of the session where users described the system by its role.}
    \centering
    {
    \caption{Users' mental model of the system before and after familiarization (percentage of the session where users described the system by its role, role names adapted from~\cite{Ross})}
    \label{tab:toolComparison}
    \begin{tabular}{cccc}
        \toprule
        Description of the system & Before fam. & After fam. & Change \\
        \midrule
        Tool & 92.98\% & 85.71\% & -7.27\% \\
        Content Generator & 77.19\% & 88.89\% & 11.70\% \\
        Co-pilot & 40.35\% & 46.03\% & 5.68\% \\
        Reference Guide & 38.60\% & 36.51\% & -2.09\% \\
        Collaborator & 29.82\% & 41.27\% & 11.45\% \\
        \bottomrule
        
    \end{tabular}
    }
\end{table}

\subsection*{Ownership and Involvement}

After every interaction with the system, we asked users their degree of ownership over the hook\footnote{Survey question: Please rate the following statement from 1 - 7 (Strongly Disagree - Strongly Agree): "In this round, I feel a strong sense of ownership over the final hook I wrote, reflecting my personal style, effort, and creativity."}.  On average, users perceive a 13.3\% increase in ownership over hooks \textit{after} familiarization (p-value < 0.005). Before familiarization, users had an average ownership perception of 4.28/7, showing their attitude towards ownership was neutral. After familiarization, their ownership perception increased to 5.21/7, showing they were between somewhat agree and agree that they felt strong ownership over their hooks. One reason the user P11 gave was that, over time, they put more effort into using the system and that their involvement contributed to increased feelings of ownership --- from simple efforts like more carefully reading and selecting LLM outputs to making more edits to both the text and the prompts.

\vspace{2px}
\begin{quote}
\begin{quote}
\textit{"Now thinking back, I would say I feel my ownership is increased... 
One [reason] is that I actually read through every sentence and every word to make sure I feel good about it...
So just by reading through all of these [system generations and...] choosing the one that I agree with make me feel like I own it... Over the 3 weeks, I started to rewrite more of the hooks, and that actually makes me feel even stronger with my ownership."
} 
\begin{flushright}
    --- P11, Week 3 Interview
\end{flushright}
\end{quote}
\end{quote}

\vspace{10px}

To investigate the relationship between involvement and ownership, we looked at our interview data. During our week three interviews, we asked the users, "How did your involvement within the sessions change after familiarization?" and "Do you find your level of involvement influences your sense of ownership?"
From the interviews, there were four types of answers stemming from different experiences with the system:






\subsubsection*{\textbf{1. More involved, more ownership: "My hard work pays off" (4 users)}} 
P6, P10, P11, and P12 shared that they became more involved in the process after familiarization. All of them mentioned that after gaining a better understanding of the system, they felt confident and comfortable enough to modify the prompt, thus dedicating more involvement and effort to the session. Specifically, P6, P10, and P12 iterated on their prompt edits multiple times to find a favorable output aligned with their interests. P11 and P12 focused on reading every result closely to select the best option. Additionally, P10, P11, and P12 frequently edited the outputs and conducted fact-checks through web searches to verify the information's accuracy. Thus, throughout this process, they all shared an increased sense of connection and ownership to the generated artifacts. 

\vspace{3px}
\begin{quote}
\begin{quote}
\textit{"After making more edits and being more engaged in the interaction, I feel a stronger connection to the tone and effort."
} \begin{flushright}
    --- P10, Week 3 Interview
\end{flushright}

\end{quote}
\end{quote}
\vspace{10px}


\subsubsection*{\textbf{2. Less involved, but maintain ownership. "I still lead the system" (2 users)}}
P7 and P8 found themselves less involved over time but still maintained high ownership. They found the system reliable and capable enough to assist with the task after observing its commendable performance over a span of three weeks. Thus, they started to allow the system to take on more work and made fewer but more concise edits. They shared that they still "\textit{led the system}" (P8) and their guidance over the system still makes a huge impact on the final artifact. 

\vspace{3px}
\begin{quote}
\begin{quote}
\textit{"I feel like, even if I did no revision at all, I'd still feel a high ownership because I did all the selections [and] I did guide the system through... if I have specific revisions in the words or if I have a revision in the prompts, I may feel more ownership over it versus if I did nothing... but still it's overall pretty high ownership generally."}
\begin{flushright}--- P8, Week 3 Interview
\end{flushright}
\end{quote}
\end{quote}
\vspace{5px}

\subsubsection*{\textbf{3. Less involved, less ownership. "I am now just a reviewer" (4 users)}}
P1, P3, P4, and P9 found themselves less involved with a drop in perceived ownership. Similar to P7 and P8, these four users also found the system reliable and capable. Moreover, P3 and P4 noted that they trusted the system more over time. Consequently, they started only providing high-level feedback to the system and reviewing its generation. For example, P3 expressed that, over usage, they increasingly allowed the system to handle more parts of the tasks. P3 mentioned they now acted more as reviewers:

\vspace{2px}
\begin{quote}
\begin{quote}
\textit{"I know the system's capability better... I can trust its ability to generate some everyday examples and a final hook... I have become more reliant on the system... I become more like a human reviewer to help guide the GPT to generate a better hook... [so now] I think my ownership dropped a lot..."}
\begin{flushright}
    --- P3, Week 3 Interview
\end{flushright}
\end{quote}
\end{quote}

\vspace{8px}

\subsubsection*{\textbf{4. No changes in involvement/ownership over time. "Topic dictates effort" (2 users)}}
P2 and P13 didn't report changes in their involvement or perception of ownership. They both found their experience was predominantly influenced by how well the LLM was able to support their chosen topic. P2 mentioned that the involvement increased only when they found the system failed to support their hook topics. P13 mentioned that they cannot fully trust the LLMs and have to intervene along the way. But they both agreed that the more edits they made, the more ownership they felt, but it wasn't a function of time. P13 compared their system experiences to mentoring undergraduate interns:
\vspace{3px}
\begin{quote}
\begin{quote}
\textit{"It's like an undergrad intern... I had to teach [the system about my topic]. I couldn't fully trust GPT... [It's like] working with my undergrad, and I tell him what to code [...but not] how to code…
Then he comes back to me, and we go over [the code to see...] if there's anything that needs more rigorous thinking... what's wrong... and how should we deal with it."
}
\hspace{55px}--- P13, Week 3 Interview

\end{quote}
\end{quote}

\vspace{11px}

In conclusion, although average feelings of ownership increased, there were nuanced ways that users experienced ownership after multiple experiences with the system. There does seem to be a relationship between effort and ownership. Based on the current state of LLM technology, many users find it necessary to edit text and prompts at every stage, but it is also a function of the topic - some topics are easier for LLMs to get right.  


\newpage
\section{Discussion}

\subsection{Why Does the Novelty Not Wear Off?}
This paper presents a three-week study of an LLM-based workflow to support science writing. Though generative AI has impressive abilities at first glance, it remains unclear whether those abilities stem from the initial impression of a new technology or from the genuine utility of the system's workflow. This study found that there was a familiarization phase (lasting 4.27 sessions on average) wherein users were exploring the system and its capabilities. After familiarization, they no longer explored its novel capabilities but focused solely on performing the task. They understood the workflow better, had a better ability to anticipate system generation, and performed better in writing hooks. Surprisingly, we found the average system utility \textit{increased} by 12.1\% after familiarization (5.47/7 to 6.32/7, p-value < 0.005), indicating that not only that the novelty does not wear off, but the workflow becomes more useful over time. We discuss potential reasons that the novelty did not wear off. 

\subsubsection*{\textbf{Reason \#1: The workflow is supporting a task with cognitively challenging steps.}} 
One possible reason the workflow maintains its usefulness over time is that the underlying task targeted by the AI workflow has steps with a high cognitive load. A formative study with scientists writing Tweetorials confirmed it was a cognitively challenging task
~\cite{Tweetorial_CSCW}. When analyzing the cognitive load through the cognitive process theory of writing~\cite{FlowerHayes1981}, they found there were two mentally demanding steps that writers wanted help with: \textit{ideation} and \textit{translation} (drafting complex thoughts into a linear narrative). Also, these steps generally do not get easier over a short period of time, even with a supporting system. 



The Tweetorial Hook workflow supports \textit{ideation} in steps 1, 2, and 3. Together, these help users identify a specific and relatable use case that will be the vehicle for hooking non-technical audiences. Thus, such AI workflows can help brainstorm multiple options to help users overcome fixation and increase ideas' number, diversity, and quality \cite{popblends,opal,3dalle,anglekindling,visar,glassman_elephant}. 

The Tweetorial Hook workflow supports \textit{translating} the use case into prose that is jargon-free, relatable, and sparks curiosity in steps 4, 5, 6, and 7. Translation is difficult because it forces users to synthesize many pieces of information into one coherent, linear text. Often, the shorter the text, the harder it is to write, but generative AI can help produce examples of linear prose that satisfy some or all of these design goals. Thus, it can be an excellent starting point or motivation to overcome the mental activation energy needed to start drafting. Even when LLMs make mistakes, such AI workflows that produce example text help users overcome the cognitive barrier of translation~\cite{visar,ReelFramer,mirowski2022cowriting,li_teach_2023,schick_peer_2022,yang_re3_2022,petridis_constitutionmaker_2023,10.1145/3584931.3607492}.


\textbf{In general, we expect that AI workflows that prove to be useful in a single session will maintain their usefulness over time if they assist with fundamentally challenging tasks.} This includes cognitive challenges like \textit{ideation}, sensemaking challenges like \textit{clustering} and \textit{synthesis}, socially challenging tasks like \textit{providing multiple user perspectives}, and tedious or time-consuming tasks like \textit{data formatting and processing}. 
    \textbf{Moreover, the utility of AI workflows can increase over time as users become more efficient at using it, better at applying it to their domain, and can easily customize it to their needs.}


\subsubsection*{\textbf{Reason \#2: Workflows improve human-AI collaboration and mitigate the effect of AI errors.}} 
Another possible reason that the workflow maintained its usefulness over time is that the task decomposition and planning provided by the workflow made it easier for users to solve the problem and to catch and correct errors made by the AI before they propagated.
Often, decomposing a hard prompt into subtasks makes it easier to solve. This has been shown for an individual doing a task~\cite{bhattacharjee2023understanding}, for groups of people doing a task~\cite{turkit,soylent,Retelny2014FlashTeams}  and for LLMs doing tasks without humans~\cite{chain_of_thought,wu2023autogen,shen2023hugginggpt}. 
Additionally, planning is also a mentally difficult process. 
Writers struggle to mentally shift between planning tasks and execution tasks, especially in writing~\cite{FlowerHayes1981}. 
Therefore, workflows that provide a plan and help people avoid task switching between high-level planning and low-level execution tasks are poised to bring lasting value.

Because workflows provide planning and task decomposition, it is fitting that workflows can enable better human-AI collaboration.
In this AI workflow and many others~\cite{visar, ReelFramer,anglekindling,opal}, AI is used to produce ideas, suggestions, drafts, and prototypes, but users control the outcome at every stage by judging the outputs and editing, regeneration, or rewriting when the AI outputs are poor. 
Generative AI is known to make many mistakes. And there are tasks that generative AI does not know how to do. 
Thus, human judgment at every step ensures that AI errors in the early steps do not propagate into bigger problems that are then harder to fix later. 
The novelty of a workflow is unlikely to wear off if it genuinely supports the problem-solving process for the human. AI suggestions for each of the steps help the user. Since workflows limit the damage of AI errors and make them possible to correct before they propagate, users are likely to keep using the system rather than abandon it due to AI errors.

%



In this study, users showed a strong preference for using the workflow rather than trying to correct the hooks fully written by AI.
In step 1 of the Tweetorial Hook workflow, there is a “quick hook” where the LLM generates a hook for the topic in one prompt (without the workflow steps).
Users were allowed to use the quick hook as their final hook and conclude the writing session if they wanted to. However, users rarely took it: out of the 120 sessions, users only used the quick hook 13 times. 
Most often, it was either factually incorrect or poorly constructed. Thus, many users expressed willingness to invest more effort in utilizing the entire workflow rather than try to fix a problematic quick hook. For example, one quick hook for the topic of ``\textit{AJAX}'' based its hook on an incorrect usage of the technology. With an incorrect example at the core of the hook, it was impossible for the user to make simple edits to correct it; it is easier to start over and use a workflow, where users could select correct usages early and build toward a hook together with the AI.  Additionally, by breaking the task into an AI workflow, there are more opportunities to customize the system at individual steps, for example, producing 20 user scenarios rather than 5, or specifying the common experiences should be relatable to a teenage audience. 

\textbf{In general, we expect that AI workflows will retain their value when the task decomposition offers problem-solving benefits to human users (even without AI). Moreover, the workflow enables better human-AI collaboration by allowing users to check and edit AI outputs to control the quality of the output at every step.}

\subsection{Customization and the Potential of Appropriation in AI Workflows}
In this longitudinal study, we found a significant increase in system utility due to a simple customization feature of exposing the prompt box so users could edit it. We believe this is a simple and powerful technology that all AI workflows should explore to help users mold the system to their needs. 
First, the prompt box helps users understand and explore the human-AI workflow rather than turning it into a ``double black box'' or hiding the affordances behind it~\cite{10.1145/108844.108867}. The prompt box facilitates transparency and offers a direct mental model of how the workflow operates in the background, enhancing user control and potential error recovery. 
Moreover, AI poses unique opportunities for end-user customization within a system. Traditional workflows are often brittle, constraining, and difficult for end users to edit~\cite{workflowbrittle}. However, with natural language editing, users can customize workflows in their own words. 

Beyond customization, there is the potential for users to appropriate a workflow for tasks it was not originally designed for. In the seminal paper on Designing for Appropriation~\cite{DesigningForAppropriationAlanDix}, Alan Dix summarizes 7 guidelines for building systems that ``\textit{plan for the expected}''. In the age of generative AI, many of these principles have become trivial to apply to systems. Systems can ``\textit{provide visibility}'' by exposing prompts; they can ``\textit{support not control}'' by allowing editing. The very nature of prompting with natural language ``\textit{allows for interpretation (Do not make everything in the system or product have a fixed meaning, but include elements where users can add their own meanings)}.'' If the system logs users' prompt edits, it becomes possible to ``\textit{learn from users' appropriations}''  at an astonishing speed. At an extreme, it presents an opportunity for a continuous co-design process by sharing successful appropriations with the user community.

In our interviews, users saw the potential to appropriate the system. 
For example, P2 initially encountered difficulties when they found that their theoretical topics (e.g., ``\textit{combinatorial discrepancy theory,}" ``\textit{price of anarchy}'', etc.) did not align well with the existing Tweetorial Hook workflow, which primarily focused on everyday use cases and shared user experiences. Thus, starting from session 4, they began to appropriate the system by using analogies rather than use cases to make theoretical topics relatable. After some prompt engineering, the metaphor strategy started to work. They also shared that this could be a valuable approach for other theoretical topics that lack relatable use cases, as well as for future workflow users coming from similar theoretical backgrounds.
Moreover, towards the end of the study, seven users shared expectations to appropriate the workflow to support other science communication tasks, like writing and revising their own research papers (P1, P8, P9, P10, P12, P13), drafting an elevator pitch about their research (P10), or making learning materials for themselves (P11). These are all interesting directions for this workflow to go in, and we are excited to see whether users can take it there on their own.

\textbf{In general, future human-AI workflows should provide editable prompt boxes so users can customize workflows using natural language. This facilitates user understanding and exploration of the workflow, supports diverse user needs, and has the potential to allow users to appropriate the workflow for tasks that it was not originally designed to support. } 

\section{Limitations and Future Work}

Our study focused on a specific AI workflow utilizing a particular generative model. While we believe that this choice provides insights that can be generalized to other workflows, some findings are particular to the system studied here. The familiarization period in this study lasted 4.27 sessions on average. This is particular to the Tweetorial Hook system. A workflow that was more complex or a task that was more unfamiliar might require more time to explore. Particularly, workflows that are less linear and have more pathways will take more time for users to explore. 

The Tweetorial Hook workflow uses GPT-4, but future LLMs might have better capabilities. If LLMs improve, will workflows still be needed? If LLMs become more factually correct, users may not need workflows to identify correct applications of a technology -- they might be able to rely on LLMs for that. However, even if LLMs improve on factuality, they might not improve in their ability to know what examples will resonate with a particular audience, so human judgment in the workflow will likely be important even if LLM models improve.



Our study encompassed three weeks and 11 of 12 users reported they already reached familiarization as they fully understood the technology. However, it is if users continued to use the workflow, their behavior would continue to change.
First, it is possible that some aspects of the system would be less novel if the user starts to notice more patterns in the LLM responses. However, even if the patterns become more clear, the system outputs could still be useful as a creative springboard. Second, we would hope to see more customization over time. As users repeat tasks, there are increased benefits to customization. Even a finalized workflow can continue to be iterated on for increased efficiency. 
We think there are many research questions that could be answered with more longitudinal studies of AI-based systems.

This study did not investigate the dangers of users potentially becoming dependent on an AI tool. Future studies could look at human performance on a task after the AI tools they used were taken away. This is important because some tools that students are allowed to use in academic settings might not be possible to use in corporate jobs due to data security concerns with AI systems. AI might be deskilling workers, but by providing scaffolding to complete a task, it also might help workers achieve skills. This deserves further study. 



Our study relied largely on self-reporting to identify the familiarization period. 
The novelty effect is primarily a cognitive phenomenon associated with liking or pleasure, and thus, some amount of self-assessment is common to use ~\cite{novelty_selfdetermination}.
However, more objective metrics might be applied in addition to self-assessment: measuring time spent on a task, measuring task performance (assessed through external evaluation), and measuring task competency through quizzes. Additionally, a study in the wild could use continued use or abandonment as a signal of a novelty effect. 


Our participant pool was limited to Computer Science PhD students, a demographic proficient with technology. 
Even though they possess differing levels of experience with using generative AI, their relatively high technology literacy and openness to accepting new technology might differ from that of the general population. Specifically, our user group might have already encountered AI risks and harms in various contexts before the study. This familiarity can breed trust and contribute to higher tolerance levels, potentially leading to the oversight or disregard of associated harms observed in the study. Additionally, users were not required to post their hooks publicly, which could have influenced their interaction with the system. If the stakes were higher, their critique of the system and session experiences might have been more stringent, or they might have invested more effort. Future studies might consider diversifying the user base or altering the stakes to observe any shifts in results.

\section{Broader Impact}
First, our study is among the earliest to examine the novelty effects and long-term utility of AI workflows. The study results shed light on insights for the future development of AI models and workflows based on them. The study method also contributes a methodological foundation for future long-term studies. 
Second, many of the insights align well with existing design theories, such as technology domestication theory, technology acceptance model, and human-AI interaction guidelines, thereby offering new insights for designing longitudinal interactions in the era of generative AI. 
Third, our paper highlights the potential of technology appropriation, which allows and encourages users to customize and adapt systems for new use cases. We initiate a discussion about using generative AI as design material to promote user co-design for system development. However, we acknowledge that our study lacks emphasis on the potential risks associated with such long-term interactions. 





\section{Conclusion}

To understand the novelty and long-term utility of AI workflows, we conducted a three-week longitudinal study with 12 users to understand the customization of generative AI workflow for science communication.
Our study revealed that the familiarization phase lasts for 4.27 sessions, which equates to 1.42 hours or one and a half weeks of usage. Before familiarization, users are mainly exploring the capabilities of the workflow and which aspects they find useful. After familiarization, the perceived utility of the system is rated 12.1\% higher than before, indicating that the perceived utility of AI is not just a novelty effect. The increase in benefits mainly comes from end-users’ ability to customize prompts,
and thus appropriate the system to their own needs, ranging from addressing system problems to adapting for new use cases. Though the mental models remain over time, users begin to change their level of involvement while collaborating with the AI workflow, thus leading to a shift in their perception of ownership.
This points to a future in which generative AI systems and workflows can facilitate user customization and appropriation, as well as a continuous co-design process between system developers and users.

\begin{acks}
This work is supported by NSF IIS-2202578 and NSF IIS-2129020. We thank Dorothy Zhang, Karthik Sreedhar, and Weirui Peng for their help in setting up the study. We thank Sitong Wang, Vivian Liu, Soya Park, Chinmay Kulkarni, the Interactive Creativity Lab at Adobe Research, and the anonymous reviewers for their feedback.
\end{acks}

\bibliographystyle{ACM-Reference-Format}
\bibliography{bib}

\appendix

\onecolumn

\section{Complete System Walkthrough}
\label{app}

\begin{multicols}{2}
\hspace{-30px}
\includegraphics[width=1.1\linewidth]{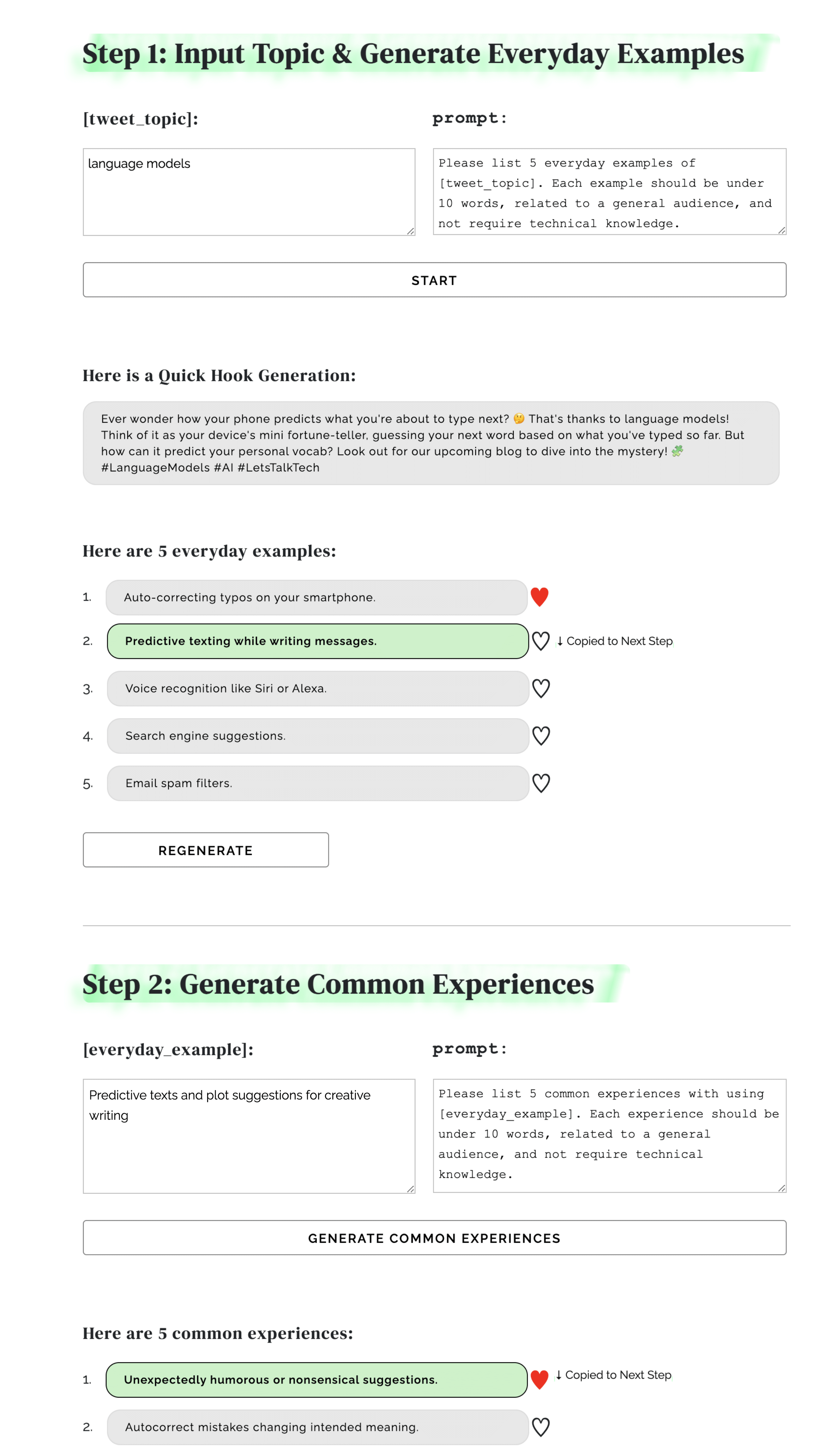}

\hspace{-30px}
\includegraphics[width=1.1\linewidth]{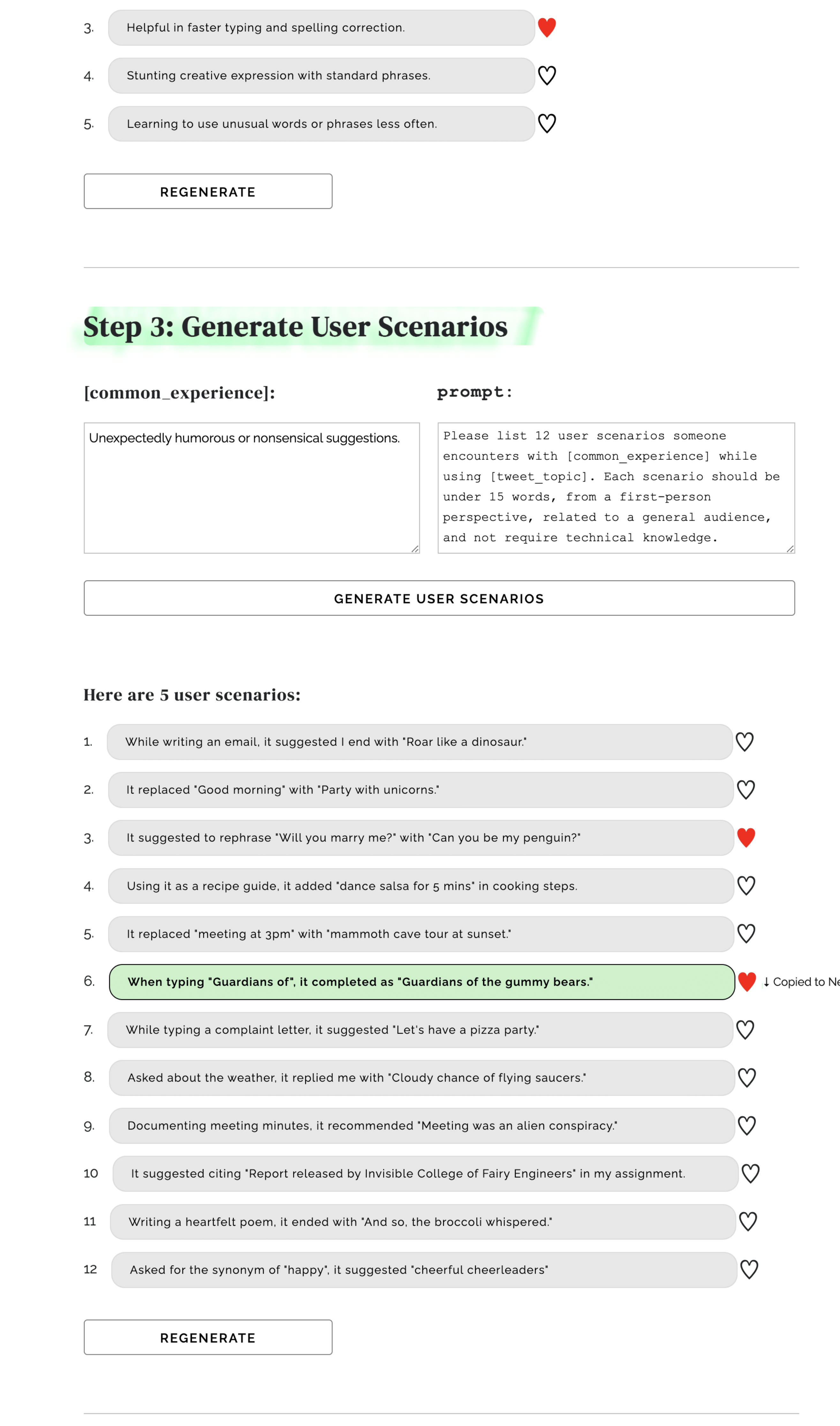}

\hspace{-30px}
\includegraphics[width=1.1\linewidth]{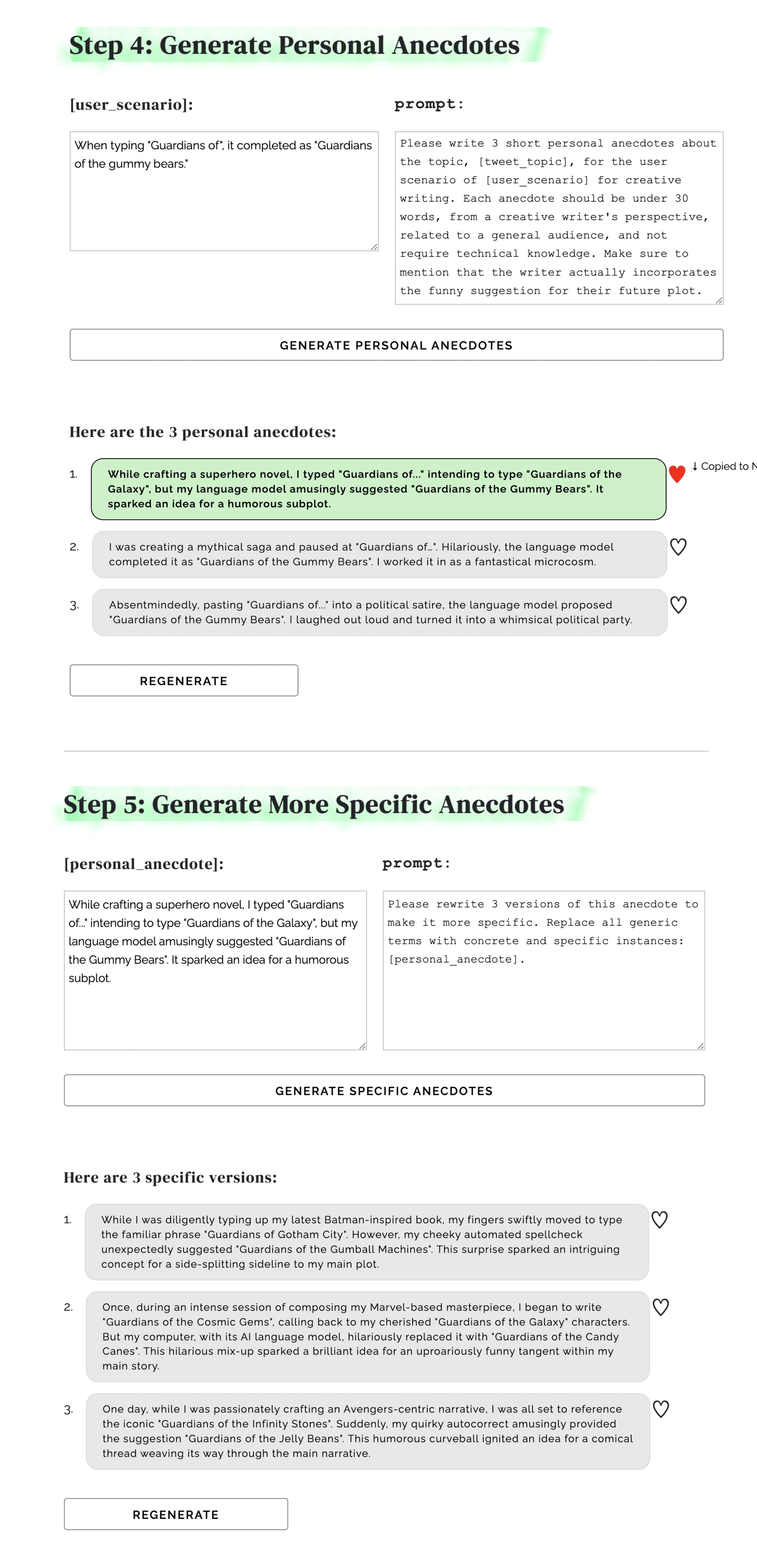}

\hspace{-30px}
\includegraphics[width=1.1\linewidth]{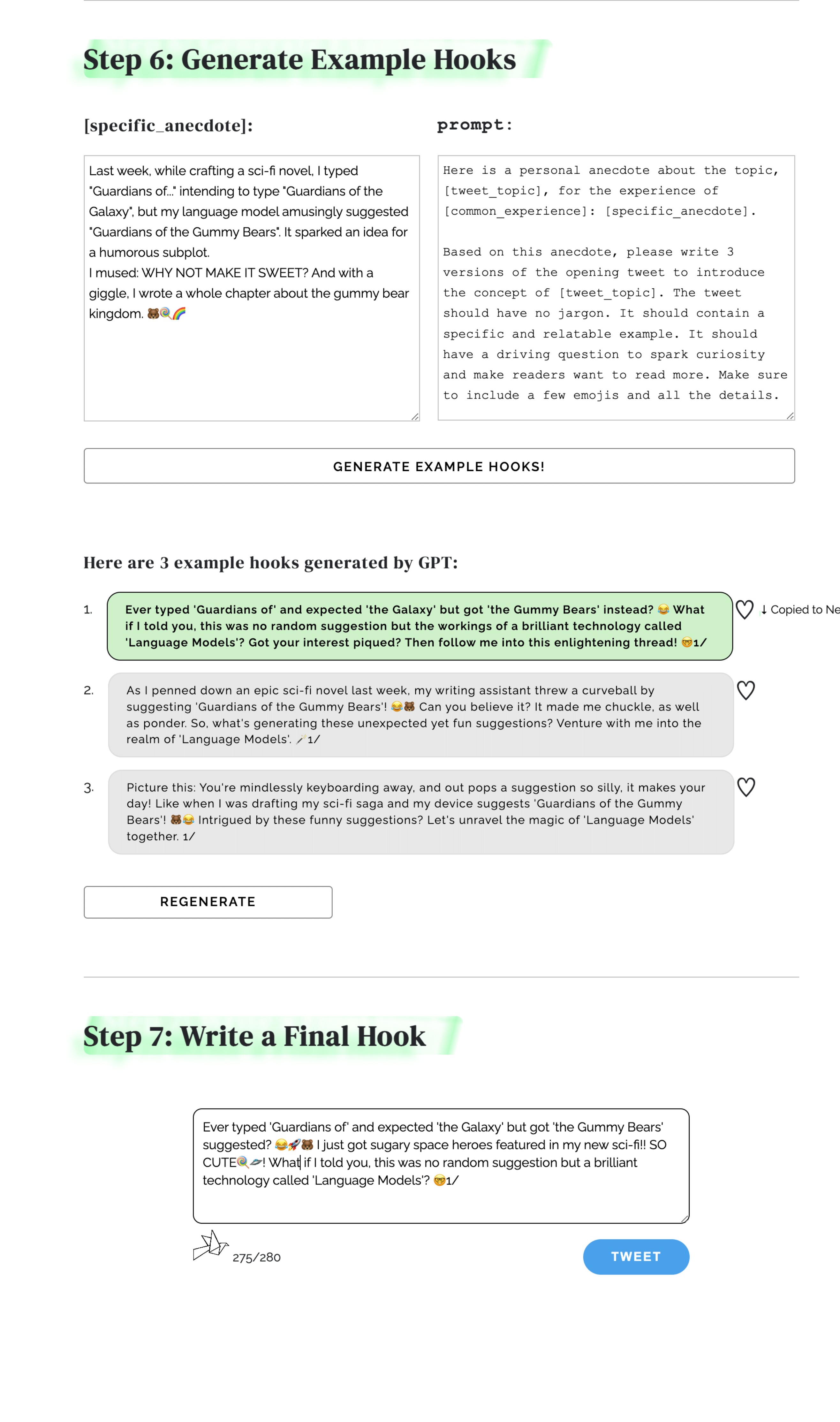}

\end{multicols}
\onecolumn
\vspace{-10px}
\section{Week 3 System Updates: Step 6 Learn-My-Style Prompt}
\label{app1}
\includegraphics[width=.78\linewidth]{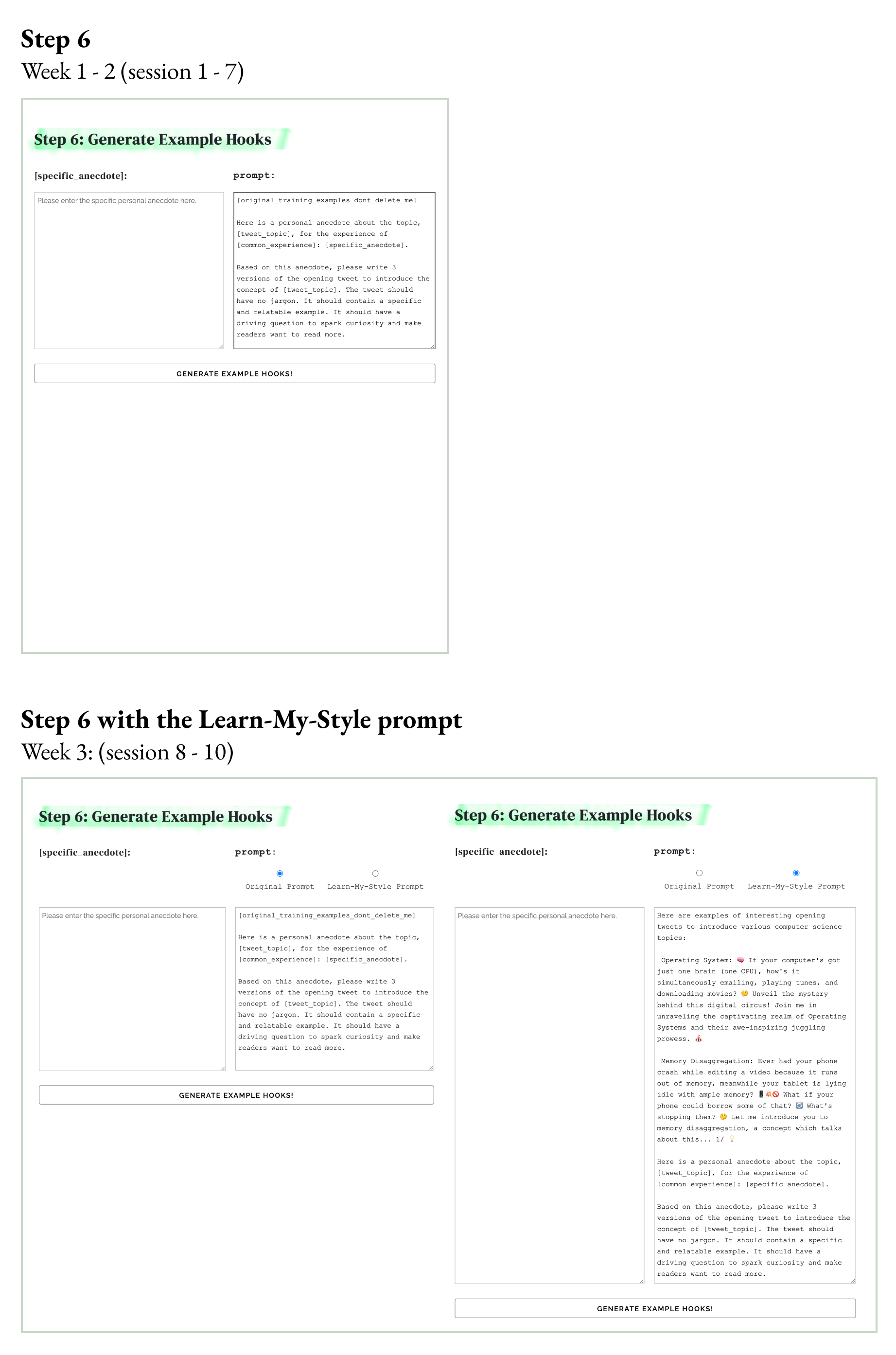}
\vspace{-60px}

\end{document}